\documentclass{article}

\usepackage[english]{babel}
\usepackage[letterpaper,top=2cm,bottom=2cm,left=3cm,right=3cm,marginparwidth=1.75cm]{geometry}

\usepackage{amsmath}
\usepackage{graphicx}
\usepackage[numbers]{natbib}

\usepackage{amssymb}
\usepackage{amsfonts,mathtools}

\usepackage[colorlinks=true, allcolors=blue]{hyperref}

\usepackage[whole]{bxcjkjatype}

\title{Collective Predictive Coding as Model of Science:\\ Formalizing Scientific Activities Towards Generative Science}

\author{Tadahiro Taniguchi$^{1,2}$, Shiro Takagi$^3$, \\Jun Otsuka$^{4,5,6}$, Yusuke Hayashi$^7$, Hiro Taiyo Hamada$^{8, 9}$}

\date{}

\begin{document}
\maketitle
\vspace{-10mm}

\begin{center}
\noindent\textsuperscript{1}Graduate School of Informatics, Kyoto University, taniguchi@i.kyoto-u.ac.jp \\
\noindent\textsuperscript{2}Research Organization of Science and Technology, Ritsumeikan University, taniguchi@em.ci.ritsumei.ac.jp\\
\noindent\textsuperscript{3}Independent Researcher, takagi4646@gmail.com \\
\noindent\textsuperscript{4}Department of Philosophy, Kyoto University, jotsuka@bun.kyoto-u.ac.jp \\
\noindent\textsuperscript{5}Data Science and AI Innovation Research Promotion Center, Shiga University, \\
\noindent\textsuperscript{6}Center for Advanced Intelligence Project, RIKEN, \\
\noindent\textsuperscript{7}AI Alignment Network, hayashi@aialign.net \\
\noindent\textsuperscript{8}DeSci Tokyo, desci.tokyo@gmail.com\\
\noindent\textsuperscript{9}ARAYA Inc., hamada\_h@araya.org\\
\end{center}

\begin{abstract}
This paper proposes a new conceptual framework called \textit{Collective Predictive Coding as a Model of Science (CPC-MS)} to formalize and understand scientific activities. Building on the idea of collective predictive coding originally developed to explain symbol emergence, CPC-MS models science as a decentralized Bayesian inference process carried out by a community of agents. The framework describes how individual scientists' partial observations and internal representations are integrated through communication and peer review to produce shared external scientific knowledge. Key aspects of scientific practice like experimentation, hypothesis formation, theory development, and paradigm shifts are mapped onto components of the probabilistic graphical model. This paper discusses how CPC-MS provides insights into issues like social objectivity in science, scientific progress, and the potential impacts of AI on research. The generative view of science offers a unified way to analyze scientific activities and could inform efforts to automate aspects of the scientific process. Overall, CPC-MS aims to provide an intuitive yet formal model of science as a collective cognitive activity.
\end{abstract}

\section{Introduction}\label{sec:1}
Science is a massive cooperative venture of mankind. Even though each person has limited observations of natural phenomena, their own experience, and perceptions and actions, we can integrate our knowledge through scientific communication and accumulate scientific knowledge throughout the global human community.  
Science is a collective behavior of humans which explore the world in an active manner. 
Meanwhile, science has its specific way of communication involving proposing hypotheses, formulating theories, conducting experiments, writing papers, submitting them to journals and conferences, and performing peer reviews.
While they have diversity depending on the specific scientific field, most people believe that such scientific activities integrate our knowledge and lead us to better understand the world.
However, in what sense do the overall scientific activities bring us a model and better understand the world? How can we improve the rules, regulations, games, and habits in our scientific activities? 
To understand and improve scientific activity itself, a wide range of studies have been undertaken, including philosophy of science, and science, technology and society (STS), and science of science~\cite{Popper1959-uu, Kuhn1962-wu, Lakatos1978-fw, latour1979laboratory,Fortunato2018SciSci}.
Despite these efforts, we have not obtained an intuitive framework that models the total scientific activities. Considering the recent changes in the scientific field, e.g., the introduction of artificial intelligence (AI) technologies~\cite{wang2023scientific} and huge problems of research ethics in science including the replicability crisis~\cite{baker20161} it is highly important for us to have a systemic view of science as a whole. This paper proposes a new conceptual and computational framework of science based on the idea of \textit{collective predictive coding (CPC)}~\cite{taniguchi2024collective}.

\begin{figure}[tb]
\centering
\includegraphics[width=0.9\linewidth]{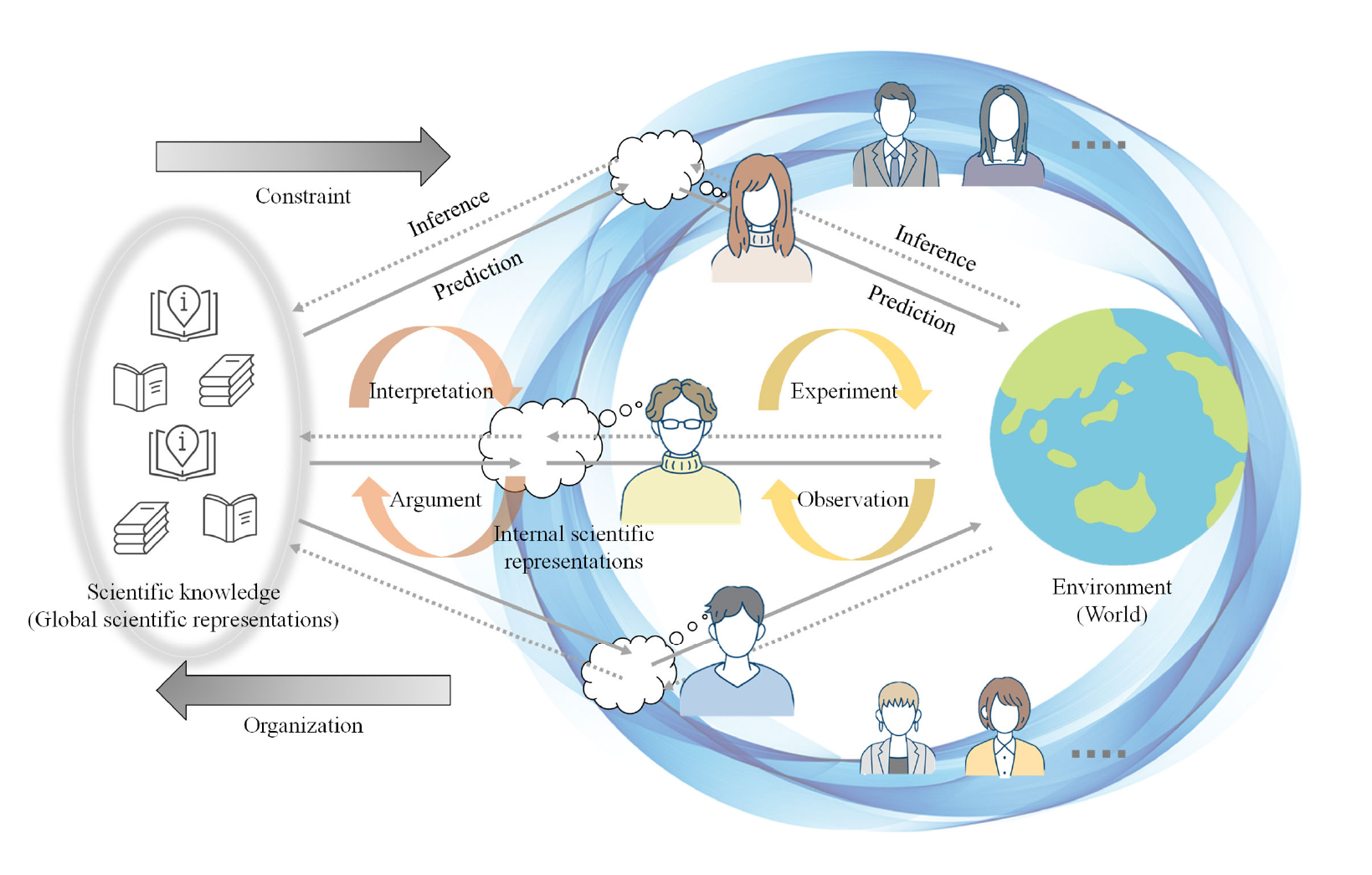}
\caption{\textbf{Overview of Collective Predictive Coding as Model of Science}. This figure illustrates how scientific knowledge emerges through collective processes. Scientists conduct experiments and observations, engaging with the target objects, i.e., environment, through their research methodologies. These individual interactions generate data and insights, which are then synthesized and encoded into a shared system of scientific knowledge, analogous to how language emerges in linguistic systems.  The diagram depicts a cyclical nature of this process, where internal scientific activities contribute to and are informed by the collective understanding, forming a dynamic feedback loop between personal research and the broader scientific corpus.}
\label{fig:cpc}
\end{figure}

The CPC hypothesis was originally proposed to explain the phenomenon of symbol emergence~\cite{taniguchi2024collective,taniguchi2016symbol,taniguchi2018symbol}. 
Symbol emergence refers to the process by which a shared system of symbols, such as language, arises and evolves within a population of agents through their interactions with each other and the environment. 
The CPC hypothesis extends the idea of \textit{predictive coding (PC)} and \textit{free-energy principle (FEP)}~\cite{friston2010free,hohwy2013predictive,friston2019free,friston2021world,taniguchi2023world} to a group of people and argues that we, humans, improve prediction capability by updating the symbol system, i.e., external representation systems, we have as a group.
In cognitive science and related fields, PC, which can be generalized from a probabilistic viewpoint to the FEP, has become a dominant idea, which regards human cognition as a process of reducing prediction errors~\cite{friston2010free,clark2013whatever}. 
The PC and FEP provide the generative view of cognition. 
FEP and PC address not only perceptual aspects of cognition but also operational aspects meaning decision making and active exploration, i.e., actions. However, basically, the idea of PC and FEP are about individual cognitive processes. In contrast, the CPC hypothesis was proposed for symbol emergence, considering language as a representative target symbol system. CPC is a generative model of symbol emergence in human society and provides a generative perspective on the emergence of language.
Notably, scientific knowledge we share in our society is also a representative symbol system. Therefore, the authors reached an insight that CPC can be extended to the explanatory theory for formalizing science activities (Figure~\ref{fig:cpc}). 

Science, including natural, social, and a wide range of sciences, attempts to model and understand the world. Although the quality of theories in science is evaluated in many ways and the values of theories are not monolithic, prediction capability is a central criterion to evaluate scientific theories \cite{Popper1959-uu, Lakatos1978-fw}. 
In particular, theories of physics provide generative models of phenomena. For example, Newtonian equations tell us the generative process of the trajectory of a target object~\cite{Landau1982}.
In contrast, related to PC and FEP, world models are considered to be an internal model which a robot or an AI has and updates to predict their sensory information depending on their actions~\cite{ha2018world,friston2021world,taniguchi2023world}. In contrast, scientific knowledge we have in our society is an explicit system of knowledge that enables us to predict phenomena happening in the world better. In other words, we can regard scientific knowledge as an external representations of explicit world model which is shared among the people who study science and conduct scientific research. This view allows us to extend the CPC hypothesis to scientific activities.

Building upon these perspectives, this paper aims to introduce a novel conceptual framework called \textit{Collective Predictive Coding as a
Model of Science (CPC-MS)} to formalize and elucidate scientific activities. CPC-MS is a framework that models science as a collective predictive coding activity, as the name suggests. As we find in the later part of this paper, we can find a reasonable correspondence between CPC and scientific activities including experimentation, measurement, proposing hypotheses, testing the hypotheses, writing papers, reviewing papers, and so forth.

The aforementioned view of science presents a vision to model the entirety of scientific activities as a probabilistic generative model~\cite{bishop2006pattern}.
We term this perspective \textit{generative science}.
While developing the concept of generative science in relation to CPC, we establish clear correspondences between the probabilistic generative model underlying CPC and concrete, practical scientific activities. Crucially, CPC-MS posits that shared external representations—scientific knowledge in this context—are updated through a form of peer-review process that functions as a decentralized Bayesian inference~\cite{taniguchi2024collective}.

The view of generative science proposed in this paper also aims to provide a clear view of the total scientific activities from the viewpoint of generative models and allow people who work on science automation to have a more intuitive roadmap for automated science (see Section~\ref{sec:4}).
Importantly, the theory is totally based on terms of probabilistic modeling and machine learning. This will also allow us to have a clear view of the integration of generative AIs into scientific activities. While there has been intense discussion and significant progress in automating individual research tasks, such as information retrieval \cite{hope2023computational}, paper writing \cite{liang2024mapping}, hypothesis proposition \cite{wang2023scientific}, paper review 
\cite{liang2024can}, and experiment execution \cite{king2009automation}, much work remains to be done in automating the entire social scientific process. CPC-MS may provide a new perspective on such automation.

The remainder of this paper is organized as follows. Section 2 introduces CPC as a model for scientific activities, explaining the CPC hypothesis and its application to modeling these activities. This section also includes a detailed explanation of the mathematical framework of CPC-MS, including its formulation based on the theory of active inference. Section 3 applies the CPC model to understanding social objectivity in science, scientific progress, and the shift from confirmatory to generative science. Section 4 explores the implications of CPC for the future of science, including speculations on AI's impact on scientific practices and providing guidelines for implementing automated total scientific activity. Section 5 discusses future research directions. Section 6 concludes this paper.

\section{Scientific Activities as CPC}\label{sec:2}
\subsection{CPC Hypothesis}
\begin{figure}[t]
\centering
\includegraphics[width=\linewidth]{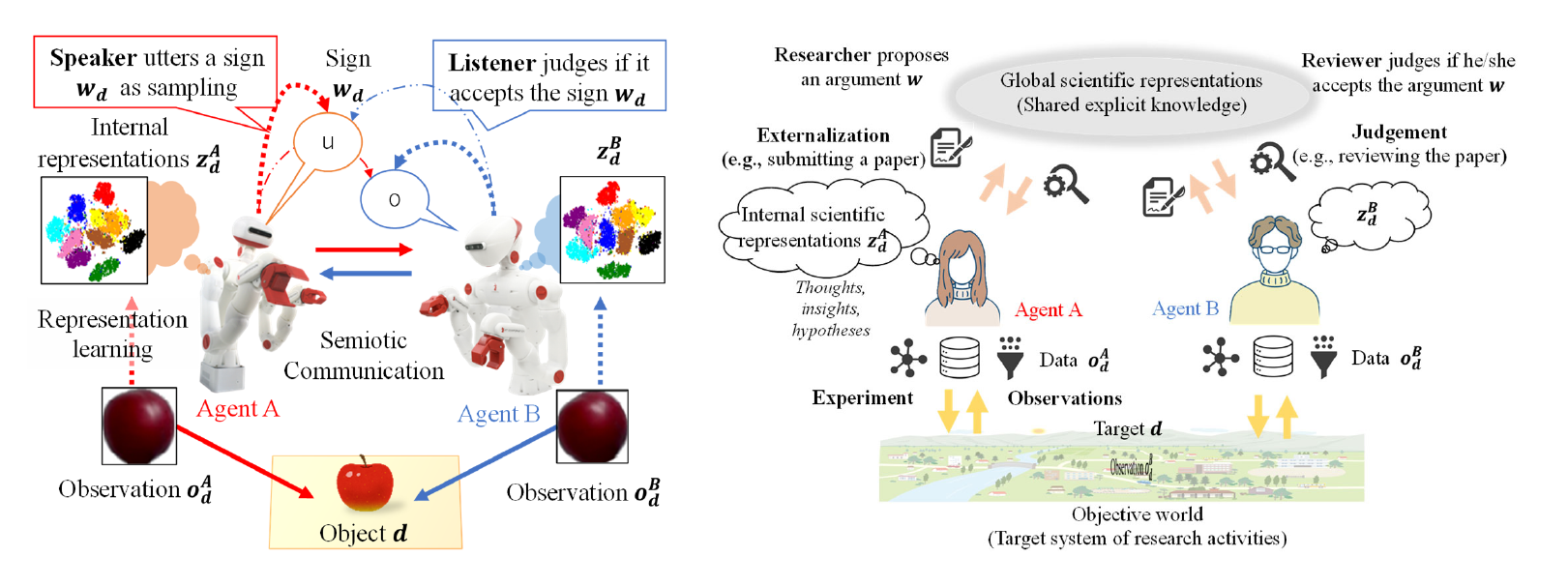}
\caption{{\bf Left}: Metropolis--Hastings naming game (MHNG)~\citep{taniguchi2022emergent}. In the MHNG, a sign $w_d$ representing the $d$-th object is regarded as a sample from an integrated posterior distribution $P(w_d| o^d_A , o^d_B)$ where $o_*^d$ is the $*$-th agent's observation. The naming game acts as a distributed Bayesian inference of the posterior distribution over the shared latent variables $w_d$ representing shared external representations. {\bf Right}: scientific activities updating shared explicit knowledge through experiments and communications involving peer-review process. The total systemic dynamics is structurally analogous to the MHNG.}
\label{fig:mhng}
\end{figure}

The CPC hypothesis proposes that the dynamics of human language can be modeled as a a process of CPC across society~\cite{taniguchi2024collective}. This hypothesis is based on the mathematical fact that a certain type of language game played among agents, i.e., Metropolis-Hastings naming game~Figure~\ref{fig:mhng}, can be acted as a decentralized Bayesian inference of a shared latent variable, sampled instantiation of which can be regarded as an utterance of a name of an object~\cite{taniguchi2022emergent}. 
The CPC hypothesis extended the idea of PC and FEP to the phenomena of language emergence in human society.
If the hypothesis is held, human language is considered to collectively encode information about the world as observed by many agents through their sensory-motor systems. 

\begin{figure}[tb]
    \centering
    \includegraphics[width=1.0\linewidth]{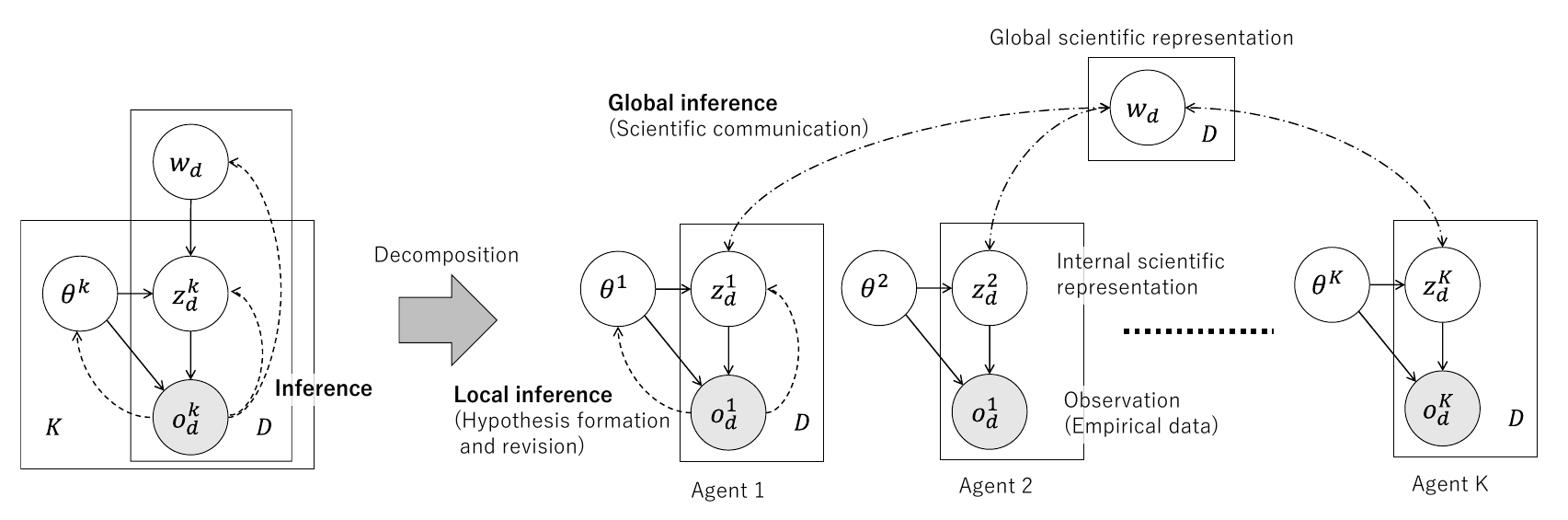}
    \caption{
Probabilistic graphical model underlying the CPC-MS framework, which is a variant of that of CPC ~\cite{taniguchi2024collective}. \textbf{Left:} An integrated model depicting the relationship between individual scientists' observations and a shared global scientific representation. In this framework, the observation $o^k_d$ of the $d$-th target by the $k$-th scientist is conceptualized as being influenced by the $d$-th global scientific representation $w_d$, mediated by the scientist's internal representation $z^k_d$. The parameter $\theta^k$ encapsulates the $k$-th scientist's learned internal models. $K$ and $D$ denote the total number of scientists and targets, respectively. \textbf{Right:} A decomposed version of the model, illustrating how the global scientific representation $w_d$ evolves through inter-scientist communication, acting as a decentralized Bayesian inference. This process is exemplified by paper submission and peer review. Through this mechanism, $w_d$ emerges as a global scientific representation of the $d$-th target.}
    \label{fig:cpc-pgm}
\end{figure}

From a mathematical perspective, CPC can be modeled using probabilistic graphical models (PGMs) where latent variables representing shared external representations are inferred in a decentralized way through agent interactions and communication. Figure~\ref{fig:cpc-pgm} represents a general form of CPC, ignoring the temporal and dynamic aspects of agent-environment sensory-motor interactions, and focusing on representation learning perspectives. 
The computational model for CPC was obtained by extending the PGM for individual representation learning to a social one. We first defined the total PGM by integrating multiple elemental modules representing agents, i.e., people. Each elemental PGM was assumed to involve latent variables $z_d^k$ representing the {\it internal representation} of an agent, and \textit{observations} $o_d^k$ corresponding to the $k$-th agent. A shared latent variable $w_d$ is placed as a parent node of $\{z_d^k \}_k$. The latent variable $w_d$ represents a sign, e.g., a sentence in language, which can be generally called an {\it external representation}. The generative model in Figure~\ref{fig:cpc-pgm} characteristically corresponds to a two-layer hierarchical representation learning system, in which internal representations $\{z_d^k \}_k$ owned by each person are influenced by and connected via a prior distribution parametrized by $w_d$. Symbol emergence corresponds to the inference of the shared external representation.
A nomenclature of variables is provided in Appendix~\ref{sec:nomenclature}.

The CPC hypothesis assumes that a language game played between agents acts as a decentralized Bayesian inference and the posterior distribution $p(w_d| \{ o_d^k\}_k)$ can be inferred approximately. Inukai et al. showed that recursive MHNG is an example of such a language game.


Mathematically, 
the inference of \( q\left(z^k \mid o^k\right) \) corresponds to representation learning by the $k$-th agent, and \( q\left(w \mid \{z^k\}_k \right) \) is assumed to be estimated through a language game within a group. As a whole, symbol emergence by a group is performed to estimate \( q\left(w, \{z^k\}_k \mid \{o^k\}_k\right) \) in a decentralized manner. We assume that we cannot estimate the true posterior $p\left(w, \{z^k\}_k \mid \{o^k\}_k\right)$, but can only estimate its approximate distribution $q\left(w, \{z^k\}_k \mid \{o^k\}_k\right)$. In MHNG, $q(w| \cdot)$ is considered to be approximated by the sample distribution $q(w| \cdot) \approx \frac{1}{I} \sum_i \delta(w, w^{[i]})$ where $\{w^{[i]}\}_i \sim p(w| \cdot)$ and $I$ is the number of samples, i.e., a Monte Carlo approximation.

However, if humans could participate in a language game that enables decentralized inference of \( w_d \), akin to the MH naming game, a symbolic system like language could develop to unify the sensorimotor information collected by individual agents. Consequently, a symbolic system emerges when agents work together in PC. A more detailed formulation from the perspective of the FEP can be found in Section~\ref{sec:2.3}.

In summary, the CPC hypothesis proposes that language emerges and evolves as a shared representation system that encodes information about the world in a compressed way, allowing agents to align their models of the world.

\subsection{CPC as a Model of Science (CPC-MS)}
\label{section:cpc-as-a-model-of-science}
The main argument of this paper is to extend the idea of CPC to a model of science. Assuming that the total goal of science is to build theories that model the world and predict future observations, the framework of CPC seems to be suitable for modeling scientific activities. Especially in science, we need to describe and share theories explicitly, even though insights and thoughts are profoundly internal phenomena in our brains. The explicit theories, which are often described in scientific papers, should be shared among society. We found a clear correspondence between language emergence and scientific activities. We call this framework \textit{CPC as a Model of Science (CPC-MS)}, as introduces in Section \ref{sec:1}.

In CPC-MS, it is assumed that each agent has partial observations and limited capability for experimentation and measurement. Based on such partial observations of the world, each agent, or scientist, tries to refine their hypothesis to predict their data more accurately. This corresponds to the representation learning of $z_d^k$. Based on such a hypothesis, an agent proposes a theory or writes a paper. The paper or theory can be accepted or rejected by reviewers. This is quite similar to the communication in MHNG.

The PGM of CPC (Figure~\ref{fig:cpc-pgm}) provides a framework for understanding and modeling the process of scientific inquiry. The key components of the PGM, including external representations ($w$), internal representations ($z$), and observations ($o$), correspond to different aspects of scientific activities. The parameter $\theta$ is a local parameter for the $k$-th agent, which represents agent-specific properties such as the agent's bias and internal world model.

We assume that scientific activity aims to reveal the structure of the world and encode it into a system of external representations. We consider a scientific paper to represent explicit knowledge that can be interpreted by everyone, unlike tacit knowledge that only the owner can access. Imagine that scientific knowledge consists of a large number of logical descriptions, e.g., mathematical equations of physical laws, structural formulas of compounds in chemistry, and descriptive theories of social systems. In this sense, $w$ represents a paper or a theoretical argument. We call $w$ \textit{global scientific representations }in this paper.

In contrast, $z$ corresponds to a personal idea about the world. The internal representation $z$ can be explicit or tacit knowledge, including thoughts, insights, and hypotheses. We call $z$ \textit{internal scientific representations} in this paper.

Table~\ref{tbl:cpc-ms} represents the correspondences between mathematical notations in CPC, phenomena in scientific activities, and language emergence. Assuming the goal of science is to obtain better external representations, i.e., theories, to better predict the world, within the framework of CPC-MS, CPC of $\{o^k_d\}_k$ and Bayesian inference of $w_d$ are the goals of science.


\begin{table}[tb]
\centering
\begin{tabular}{|c|c|c|}
\hline
\textbf{Mathematical notations}  & \textbf{Science activities} & \textbf{Language emergence} \\
\hline
 External representations $w$ & \shortstack{Global scientific representation\\ (e.g., published papers, \\established theories, \\consensus models)} & \shortstack{Shared symbolic system\\ (e.g., words, \\sentences, \\signs)} \\
\hline
Internal representations $z$   & \shortstack{Internal scientific representations\\ (e.g., hypotheses, insights, \\mental models, intuitions)} & \shortstack{Cognitive representations\\ (e.g., concepts, \\mental images,\\perceptual state)} \\
\hline
Observations $o$ & \shortstack{Empirical data\\ (e.g., experimental results, \\field observations, measurements)} & \shortstack{Sensory experiences\\ (e.g., visual, auditory, \\tactile inputs)} \\
\hline
Inference of $P(z|o)$  & \shortstack{Hypothesis formation and revision\\ (e.g., data analysis, \\theory development)} & \shortstack{Representation learning\\ (e.g., categorization, \\concept formation)} \\
\hline
Inference of $P(w|z)$ & \shortstack{Scientific communication\\ (e.g., paper writing, peer review, \\oral discussion)} & \shortstack{Language game\\ (e.g., speech production, \\interpretation)} \\
\hline
\end{tabular}
\caption{Correspondences between mathematical notations in CPC, phenomena in scientific activities and language emergence} \label{tbl:cpc-ms}
\end{table}

The following is a scenario of two-agent scientific communication, which can be extended to a multiple-agent setting, playing scientific communication possibly corresponding to a decentralized Bayesian inference in the same way as MHNG. The schematic correspondence between MHNG and the scientific communication is depicted in Figure~\ref{fig:mhng}.

\begin{enumerate}
    \item {\bf Experimentation and Measurement:} Each agent (scientist) performs experiments on the $d$-th target and collects observations, which represent the data $o=\{o^*_d\}_d$. 
    Due to limited access to the world, scientists can only perform a subset of possible experiments and obtain partial observations.
    \item {\bf Testing and Refining Hypotheses:} Based on the observations ($o=\{o^*_d\}_d$), scientists update their internal scientific representations, i.e., internal representations, ($z^*_d$) through the inference process $P(z|o)$. This process mirrors how scientists evaluate and refine hypotheses in light of experimental results. Testing hypotheses is represented by the evaluation of the likelihood term $P(o|z)$ in the CPC model, from generative perspectives. Scientists seek hypotheses that maximize the likelihood of the observed data.
    \item {\bf Externalization of Scientific Representations:} Scientists externalize their ideas mainly by writing papers to communicate their internal representations ($z$) to the broader scientific community as explicit knowledge. This process can be modeled as a sampling process from the distribution $P(w|z)$, mapping internal representations to external representations ($w$), which correspond to theories and models described in scientific papers.
    \item {\bf Judgment of Scientific Representations:} When a scientist submits a paper ($w$) for review, other scientists evaluate it based on their own internal representations ($z$). Reviewers assess the compatibility of the proposed theory with their understanding and their beliefs. 
    If successful, the distribution of accepted papers can be regarded as samples of $q(w \mid \{z^k\}_k)$, similar to the MHNG. In other words, when reviewer $k'$ makes acceptance decisions based on $p(z^{k'}|w)$, the sample distribution of $w$ is influenced by reviewer $k'$'s knowledge, resulting in a sample distribution that approximates $p(w|z^{k}, z^{k'})$ rather than $p(w|z^{k})$. This means that through the judgment process, reviewer $k'$'s knowledge is integrated into the distribution~\footnote{Note that there are two types of interpretation of the review process. One is that two researchers, the $k$-th and the $k'$-th agents, have different observations $o^k \neq o^{k'}$, and the reviewer $k'$ judges the paper $w$ written by  the $k$-th agent, i.e.,  a sample from $p(w|o^k)$, by using $q(z^{k'}|o^{k'})$. However, in this case, if there is a significant difference between the distributions $p(o^k)$ and $p(o^{k'})$, it may result in an unfairly low acceptance ratio. The other interpretation is that two researchers share the same data $o^k = o^{k'}$, i.e., the data is shared publically. In this case, the reviewer uses their knowledge, i.e., the global parameter of $p(z^{k'}|w)$ and $p(z^{k'}|o^{k'}=o^k)$, without being affected by the difference in data distributions.}.
    In other words, this scientific communication consisting of sampling, i.e., externalization, and judgment, i.e., peer review process, is considered as the approximate (decentralized) Bayesian inference of $P(w|z)$ in the CPC framework. 
    \item {\bf Iteration:} The process returns to step 1, with scientists designing new experiments based on the updated consensus, continuing the cycle of scientific inquiry.
\end{enumerate}

Through ongoing communication of theories and evidence via papers and discussions, the scientific community, here the group of two agents, updates its shared understanding of the world. This process corresponds to the decentralized Bayesian inference of shared external representations ($P(w|o)$). In generative science, the ideal scientific consensus corresponds to the inference of the posterior distribution, i.e., $p\left(w  \mid \{o^k\}_k\right) \approx q\left(w \mid \{o^k\}_k\right)$.
Notably, the CPC-MS framework models social scientific activity from the viewpoint of Bayesian inference that aggregates the evidence obtained by agents who have partial capabilities. The CPC-MS framework does not necessarily provide a method enabling the agents to reach the objective ``truth,'' but rather provides a view of scientific communication to better understand, justify, and refine the mechanism of scientific activities played by the multi-agent system, i.e., scientists.

By mapping the components of the CPC framework to specific scientific activities, we can see how the PGM provides a unified view of the scientific process. The interplay between observations, 
 internal representations, and external representations in the CPC model mirrors the way scientists gather evidence, propose hypotheses, test theories, and communicate their findings to the broader community. The CPC framework offers a principled way to understand the emergence of scientific knowledge through the collective efforts of individual scientists.

\subsection{Active Inference on CPC-MS}\label{sec:2.3}
In transitioning from a MHNG-based perspective of the CPC model to one grounded in active inference, we open a new dimension of understanding scientific processes. Active inference~\cite{parr2022activeinference}, which has been pivotal in explaining cognitive processes through the minimization of prediction errors, offers a robust framework for reinterpreting the CPC model. This approach not only aligns with the principles of active inference but also provides a comprehensive view of scientific inquiry as a collective cognitive process. By adopting this perspective, we can better elucidate how individual scientists' observations and hypotheses are integrated into a cohesive body of scientific knowledge, facilitating a deeper understanding of the dynamics of scientific progress and the role of collective intelligence in scientific discovery. This shift in perspective underscores the potential of CPC-MS to model science as an active, generative process, driven by the continuous interplay between prediction and observation within the scientific community~\cite{balzan2023distributed, taniguchi2024collective}.

Because the CPC-MS regard the science community as a collective intelligence performing a decentralized Bayesian inference from the viewpoint of generative models, the system as a whole is also viewed as an integrated cognitive system that performs free-energy miniization and active inference.  Within the system, each indivisiual is drived to explore based on their individual and collective motivations. Active inference is a theory that explains how living things understand their surroundings, act in the world, and learn. It's based on the idea that organisms try to minimize the difference between what they expect and what they actually experience, both now and in the future. Several studies have demonstrated that active inference can explain how organisms balance exploring new options with exploiting known rewards. This approach offers a clear mathematical solution to a long-standing problem in decision-making research.

Mathematically, the CPC-MS in the simplest case can be described as follows:
\begin{align}
\text{Generative model:}\quad & p\left(w, \mathbf{z}, \mathbf{o} \mid \mathbf{a}, \mathbf{C} \right) = p(w) p\left(\mathbf{o} \mid \mathbf{z}, \mathbf{a}, \mathbf{C}\right) p\left(\mathbf{z} \mid w, \mathbf{a}\right) \\
\text{Inference model:}\quad & q\left(w, \mathbf{z}, \mathbf{o} \mid \mathbf{a}, \mathbf{C} \right) = q\left(w \mid \mathbf{z}\right) q\left(\mathbf{o} \mid \mathbf{C}\right) q\left(\mathbf{z} \mid w, \mathbf{o}, \mathbf{a}\right) 
\end{align}
where $w$ is external (collective) representations, $\mathbf{z} = \{z^{k}\}_k$ is the set of representations by the $k$-th agent,  $\mathbf{o} = \{o^{k}\}_k$ is the set of observations by the $k$-th agent, $\mathbf{a} = \{a^{k}\}_k$ is the set of sequence of actions taken by the $k$-th agent, and $\mathbf{C} = \{C^{k}\}_k$ is the set of sequence of rewards taken by the $k$-th agent. The inference of \( q\left(\mathbf{z} \mid w, \mathbf{o}, \mathbf{a}\right) \) corresponds to representation learning by the $k$-th agent, and \( q\left(w \mid \mathbf{z}\right) \) is assumed to be estimated through a language game within a group. As a whole, symbol emergence by a group is performed to estimate \( q\left(w, \mathbf{z} \mid \mathbf{o}, \mathbf{a}, \mathbf{C} \right) \) in a decentralized manner. We assume that we cannot estimate the true posterior $p\left(w, \mathbf{z}, \mathbf{o} \mid \mathbf{a}, \mathbf{C}\right)$, but can only estimate its approximate distribution $q\left(w, \mathbf{z} \mid \mathbf{o}, \mathbf{a}, \mathbf{C}\right)$. 

Based on these premises, the negative variational free energy $-F$ of the CPC-MS, which exactly corresponds to the model's variational lower bound (ELBO), is defined as follows:
\begin{align}
F &= D_{\mathrm{KL}}\left[q\left(w, \mathbf{z}, \mathbf{o} \mid \mathbf{a}, \mathbf{C}\right) \| p\left(w, \mathbf{z}, \mathbf{o} \mid \mathbf{a}, \mathbf{C}\right) \right] \\
&= \int q\left(w, \mathbf{z}, \mathbf{o} \mid \mathbf{a}, \mathbf{C}\right) \ln{\frac{q\left(w, \mathbf{z}, \mathbf{o} \mid \mathbf{a}, \mathbf{C}\right)}{p\left(w, \mathbf{z}, \mathbf{o} \mid \mathbf{a}, \mathbf{C}\right)}} \mathrm{~d}w \mathrm{d}\mathbf{z} \mathrm{d}\mathbf{o} \\
&= -\mathbb{E}_{q}\left[ \ln{\frac{p(w)}{q\left(w \mid \mathbf{z}\right)}} \right] - \mathbb{E}_{q}\left[ \ln{\frac{p\left(\mathbf{o} \mid \mathbf{z}, \mathbf{a}, \mathbf{C}\right)}{q\left(\mathbf{o} \mid \mathbf{C}\right)}}\right] - \mathbb{E}_{q}\left[ \ln{\frac{p\left(\mathbf{z} \mid w, \mathbf{a}\right)}{q\left(\mathbf{z} \mid w, \mathbf{o}, \mathbf{a}\right)}} \right] \\
&\simeq \underbrace{\mathbb{E}_{q}\left[ \ln{q\left(w \mid \mathbf{z}\right)} \right]}_{\text{Collective term}} - \underbrace{ \mathbb{E}_{q}\left[ \ln{p\left(\mathbf{o} \mid \mathbf{z}, \mathbf{a}, \mathbf{C}\right)}\right] - \mathbb{E}_{q}\left[ \ln{\frac{p\left(\mathbf{z} \mid w, \mathbf{a}\right)}{q\left(\mathbf{z} \mid w, \mathbf{o}, \mathbf{a}\right)}} \right]}_{\text{Individual term}} \\
&= \underbrace{\mathbb{E}_{q}\left[ \ln{q\left(w \mid \{z^{k}\}_{k}\right)} \right]}_{\text{\text{Collective regularization}}} - \sum_k \underbrace{\mathbb{E}_{q}\left[ \ln{p\left(o^{k} \mid z^{k}, a^{k}, C^{k}\right)} \right]}_{\text{Individual prediction error}} - \sum_k \underbrace{\mathbb{E}_{q}\left[ \ln{\frac{p\left(z^{k} \mid w, a^{k}\right)}{q\left(z^{k} \mid w, o^{k}, a^{k}\right)}} \right]}_{\text{Individual regularization}}
\label{eq:free_energy}
\end{align}
where the first term $\mathbb{E}_{q}\left[ \ln{q\left(w \mid \mathbf{z}\right)} \right]$ represents the collective regularization, which corresponds to the maximization of the log-likelihood concerning the collective latent variable $w$, and the second term $-\mathbb{E}_{q}\left[\ln{p\left(\mathbf{o} \mid \mathbf{z}, \mathbf{a}, \mathbf{C}\right)}\right]$ represents the sum of the prediction error for each $k$-th agent concerning their observation $o^{k}$, and the third term $-\mathbb{E}_{q}\left[ \ln{\frac{p\left(\mathbf{z} \mid w, \mathbf{a}\right)}{q\left(\mathbf{z} \mid w, \mathbf{o}, \mathbf{a}\right)}} \right]$ represents the sum of the regularization for each $k$-th agent, indicating how useful the generation of the latent variable $z^{k}$ is for estimating the environmental system for each $k$-th agent. It's important to note that $\mathbb{E}_{q}\left[\ln{p\left(w\right)}\right]$ and $\mathbb{E}_{q}\left[\ln{q\left(\mathbf{o} \mid \mathbf{C}\right)}\right]$ are constant terms and therefore do not affect the parameter estimation in the inference model. Additionally, while individual term can be described as the sum of the values for each $k$-th agent, collective term cannot be expressed as a sum over $k$.

In Equation~\ref{eq:free_energy}, the {\it individual prediction error} term represents an individual's predictive capability based on their internal scientific representations $\left\{z^k\right\}_k$. In generative science, we take the stance of evaluating theories and hypotheses based on how well they can predict observed values, and this term corresponds to that. The {\it individual regularization} term signifies how consistent the inferred $z^k$ based on the external and explicit scientific knowledge $w$ is with the inferred $z^k$ based on one's own observation $o^k$. This term represents the interaction between top-down and bottom-up information, which is essential for PC and FEP. While the above two terms appear in general FEP, the first term, the {\it collective regularization} term, is essential for CPC. The existence of this term gives rise to language and scientific knowledge as external representation systems. This term suggests that it's better for the $w$ represented or inferred by all agents to be as consistent as possible. Notably, both $z^k$ and $w$ do not come from the environment but are created by people. The collective regularization term works to adjust latent variables. This means that there is arbitrariness here. In other words, the symbol $w$ representing $z^k$ inferred via $q(z^k| w, o^k, a^k)$ has various degrees of freedom as long as it satisfies social consensus represented by the collective regularization term. This corresponds to the ``arbitrariness,'' which is considered to be a nature of symbols in semiotics~\cite{Chandler2002}. We, humans, are adapting to the environment as a group and modeling the world in a generative manner, not only through synaptic plasticity in our nervous system but also by using semantically and structurally flexible symbol systems in society with this plasticity. CPC-MS expresses this quality of scientific theories that enables the group to be an adaptive predictor, i.e., a group of agents attempting to model the world. 

The expected free energy $G\left( \tilde{\mathbf{a}}\right)$ of the CPC-MS for future observational data is defined as follows:
\begin{align}
G(\tilde{\mathbf{a}}) &= \mathbb{E}_{q(\tilde{w}, \tilde{\mathbf{z}}, \tilde{\mathbf{o}} \mid \tilde{\mathbf{a}}, \tilde{\mathbf{C}})}[\tilde{F}] \\
&= \int q(\tilde{w}, \tilde{\mathbf{z}}, \tilde{\mathbf{o}} \mid \tilde{\mathbf{a}}, \tilde{\mathbf{C}}) \tilde{F} \mathrm{~d} \tilde{w} \mathrm{~d} \tilde{\mathbf{z}} \mathrm{~d} \tilde{\mathbf{o}} \\
&= \underbrace{\mathbb{E}_{q}\left[\ln q\left(\tilde{w} \mid\left\{\tilde{z}^k\right\}_k\right)\right]}_{\text {Collective epistemic value }} - \sum_k \underbrace{\mathbb{E}_{q}\left[\ln p\left(\tilde{o}^k \mid \tilde{z}^k, \tilde{a}^k, \tilde{C}^k\right)\right]}_{\text {Individual pragmatic value }} - \sum_k \underbrace{\mathbb{E}_{q}\left[\ln \frac{p\left(\tilde{z}^k \mid \tilde{w}, \tilde{a}^k\right)}{q\left(\tilde{z}^k \mid \tilde{w}, \tilde{o}^k, \tilde{a}^k\right)}\right]}_{\text {Individual epistemic value }}
\label{eq:expected_free_energy}
\end{align}

Scientific progress is exploratory in nature. The CPC theory provides a new perspective on collective exploration in science. Individual scientists explore research fields and reach scientific discoveries through hypothesis generation, experimentation, simulation, etc. Drivers of human exploration including scientific activities can be based on extrinsic and intrinsic motivations \cite{Berlyne1950curiosity, Kang2009IG, Lowewenstein1994IG}. Intrinsic motivation of theme selection and scientific discovery is especially described as individual drivers such as curiosity. This individual exploration of science can be supported by theories of information gain. The information gain theory states that the relationship between curiosity and confidence is a inverted U-shape function \cite{Spitzer2024IG}. Information gain as a proxy of curiosity measures the reduction of uncertainties after samplings. This information-theoretic principle inspired multiple related studies on agents' exploration \cite{Oudeyer2016LP, Schwartenbeck2019exploration}.

Analogous to the information gain theory, individual scientists attempts to seek novel scientific discovery by rewarding gaining new information based on current situation. The information gain theory of curiosity states that exploratory behaviors are driven by information gain between what an agent knows (internal representation $z^k$) and what an agent obtains (observation $w, o^k, a^k$). This process can be formulated as refinement of a hypothesis $q(z^k|w, o^k, a^k)$. However, this view lacks organizational aspects of science since theme selection of scientists are constrained by multiple collective factors such as scientific trends, grant opportunities and collaborators. In other words, individual exploration of science can be biased by global constrains. These global constrains may be interpreted as external (collective) representation $w$.

The CPC theory can capture collective aspects of scientific discovery and exploration as hypothesis updating and hypothesis testing. We here describe different exploratory patterns and scientific activities. 

\begin{itemize}
   \item {\bf Hypothesis Test:}\ In scientific activities, doing an experiment is crucial for testing hypothesis. Hypothesis testing can be driven by sampling based on researcher's internal model and theories as global representation.
    \item {\bf Hypothesis Update:}\ Updating a hypothesis is another important scientific activity. Updating an internal model of a researcher as hypothesis update is driven by the information gain between models given by observation and global representation.
    \item {\bf Theme Selection:}\ Theme selection is also exploration in scientific activities. This exploration can be driven by comparisons between information gains from multiple targets $IG_{d_1,d_2,...,d_m}$. Each scientist has preference on which target is more interesting or curious to him or her. Such preference to hypothesis update promotes theme selection of scientific topics in a decentralized manner. 
\end{itemize}



In summary, the CPC theory can describe multiple collective aspects of scientific discovery such as hypothesis test, hypothesis update, and theme selection.

\subsection{Example}
For illustrative purpose, let us explain the case of ``verifying whether drug $X$ has a pharmacological effect on disease $Y$'' within the CPC-MS framework.

\begin{enumerate}
    \item {\bf Experimentation and Measurement:} A research teams $k \in \mathbf{K}$ designs and conducts an experiment to investigate the effect of drug $X$ on disease $Y$. For example, they perform clinical trials and collect data $o^k_d$ on patients' symptom improvement and biomarker changes. The action $a^k_d$ here corresponds to the specific experimental procedures, such as administering the drug, measuring biomarkers, and recording patient outcomes.   Though multiple research teams can obtain their respective $o^k_d$ $(k \in \mathbf{K})$, each group can only perform a subset of possible experiments and obtain partial observations due to ethical constraints, experimental design limitations, and  limitations, such as constraints on experimental equipment and number of subjects.
    
    \item {\bf Testing and Refining Hypotheses:} Research team $k$ analyzes the observational data $o=\{o^k_d\}_d$ and evaluates whether drug $X$ shows a statistically significant effect on disease $Y$. Based on this, they update their hypothesis (internal representation) $z^k_d$ about the effect of drug $X$. This process is represented as approximate inference of the posterior distribution $p(z|w, a)$. For instance, they refine their hypothesis by evaluating the presence and strength of drug $X$'s effect through statistical analysis. From a generative perspective, the fitness of hypothesis to the data corresponds to evaluating the likelihood term $p(o|z, a, C)$. The reward $C^k$ in this context could represent the potential benefits of the drug, such as improved patient outcomes or reduced healthcare costs, which motivate the research. CPC-MS assumes that researchers pursue hypotheses that increase the likelihood of the observational data while being influenced by the top-down effect $p(z|w, a)$ (prior distribution) given by $w$, which represents prior knowledge in the field. This aligns with Bayesian inference as $p(z,o \mid w, a, C) \propto p(z|w, a)p(o|z, a, C)$, appropriately updating one's hypothesis under the influence of prior knowledge $w$. If the hypothesis ($z$) is supported, they may plan additional experiments to obtain more data ($o$) for further verification. If not supported, they consider the reasons and either review the experimental methods to verify the hypothesis or modify the hypothesis ($z$).

    \item {\bf Externalization of Scientific Representations:} Researchers externalize their internal scientific representations $z$ as papers and communicate them to the scientific community. This can be modeled as a sampling process from the distribution $P(w|z)$. Specifically, they write a paper $w$ on the theory and data analysis results regarding the effect of drug $X$ to convert internal representations into external representations.

    \item {\bf Judgment of Scientific Representations:} When a paper $w$ is submitted, other researchers (i.e., reviewers) evaluate it. Reviewers judge the validity of the proposed theory based on their internal representations $z$, which are informed by existing knowledge in the world and their own expertise. They evaluate aspects such as the appropriateness of the experimental design, the accuracy of statistical analysis, and the validity of conclusions.
    
    This corresponds to the Listener's acceptance judgment in MHNG, and when done properly, the distribution of accepted papers can be viewed as samples from $q(w \mid \{o_d^k\}_k)$. The scientific activities through this process, i.e., CPC-MS, can be considered as approximate distributed Bayesian inference of $p(w \mid \{o_d^k\}_k))$.

    \item {\bf Iteration:} The process returns to step 1. Based on the updated scientific consensus, researchers design new experiments. Now, other research teams $k'$ conduct experiments on the same target $d$. For example, they plan experiments to verify the effect of drug X in different patient groups or to elucidate the detailed mechanism of action, continuing the cycle of scientific inquiry.
\end{enumerate}

In this way, CPC-MS represents scientific inquiry not so much as ``the pursuit of true descriptive knowledge $w^*$'' but rather as ``an overall mechanism for integrating knowledge from limited observations.''

\section{Explaining the Scientific Activities with the CPC-MS} \label{sec:3}
This section examines the CPC-MS framework's implications for understanding the social dynamics and progress of scientific inquiry. By emphasizing the collective nature of knowledge formation, it challenges the traditional view of science as an individual, confirmatory process. Instead, the framework highlights how diversity and social interactions among scientists contribute to objective knowledge and drive scientific advancement. 
Additionally, the Bayesian perspective reframes scientific progress not as the accumulation of truths but as the continuous refinement and generation of new ideas, emphasizing the generative aspects of scientific activity.

\subsection{Social Objectivity}
\label{section:social-objectivity}
One salient feature of the CPC-MS is that scientific knowledge emerges through social interactions among individual scientists who have only partial and possibly distorted understanding of the world. 
Importantly, the establishment of a global scientific understanding does not entail the consensus among individual scientists: they may well disagree each other ($q(z^k|o) \neq q(z^l|o)$ for $k \neq l$) or have incorrect posterior ($q(z^k|o) \neq p (z^k|o)$) even at the limit where the community as a whole approximately attains the true posterior $q\left(w  \mid \{o^k\}_k\right) \approx p\left(w \mid \{o^k\}_k\right)$. 
The holders of scientific knowledge, therefore, are the scientific community as a whole rather than individual scientists.
This aligns with the social approach to scientific knowledge~\cite{Longino1990-ek, Kitcher1993-ly}, which frames the objectivity of scientific inquiry as arising from the interplay of social interactions and empirical investigation. 
In this picture, scientists are depicted not as detached truth seekers who strictly adhere to universal methodologies, but as situated individuals influenced by their backgrounds, beliefs, and values. The individual biases, however, get corrected in critical dialogues including peer reviews and replications of experimental results. 

The CPC-MS provides a Bayesian model for such social processes through which objective knowledge emerges from diversity.
As described in Section 2, scientists in the CPC-MS model may have their own parameter values $\theta$ and thus form different inner representation $z$ based on the same observation $o$. Thus, their research output $w$ well reflect individual biases. This result, however, will not be integrated into the scientific body unless it is accepted by other scientists (i.e., reviewers), who makes the decision on the basis of their own probability criterion $q(w|z)$. The key observation of the CPC-MS model is that such mutual criticisms can be seen as an essential part of the decentralized Bayesian inference, e.g., MHNG, which realizes the inference to the posterior distribution at the scientific community as a whole.

Moreover, our model predicts the ways in which diversity and heterogeneity even promote objective scientific investigations, by allowing for the inclusion of diverse perspectives and the exploration of varied research paths.
First of all, Bayesian inference becomes skewed or biased if data concentrate on a limited range, e.g., if the scientific community conducts experiments and observations only within a specific range of interests and subjects. 
Second, the asymptotic convergence to the true posterior via the repeated decentralized sampling is conditioned by the sampling process being \emph{ergodic}, which roughly means that every part of the hypothesis space is eventually explored.
This condition is violated if, among others, there is any ``blind spot'' theory $w$ that has no positive probability of being proposed by any scientist (so that the Markov chain becomes \emph{reducible}), or if scientific activities are trapped in cycles, with no chance of escaping and exploring new theoretical possibilities (so that the chain is \emph{periodic}). 
Moreover, an effective convergence of Metropolis-Hastings algorithm requires that samples are independent of each other, which, in the present context, is translated as that scientists carry out their research according to their own interests and concerns. In contrast, if particular topics or groups become too influential within a research community, scientific explorations can become skewed toward a specific region of the hypothesis space, resulting in slower convergence. 
These pitfalls may be avoided by increasing the diversity of a scientific community, ensuring a wide range of perspectives and approaches are represented.

\subsection{Scientific Progress}
\label{section:scientific-progress}
{In this subsection, we explore the crucial role of the model parameter \(\boldsymbol{\theta} = \left\{ \theta^k \right\}_k\) within our framework. As depicted in Figure 3, \(\boldsymbol{\theta}\) is a fundamental component of the Bayesian network that underpins the CPC-MS model. It serves as a key determinant of the generative and inference models employed by each agent engaged in active inference. Specifically, \(\boldsymbol{\theta}\) encapsulates the internal models learned by each agent, influencing how they interpret observations and form hypotheses about their environment.

The parameter \(\boldsymbol{\theta}\) acts as a bridge between individual internal representations and the shared global scientific representation \(w_d\). By modulating the agents' hypothesis formation and revision processes, \(\boldsymbol{\theta}\) plays a pivotal role in the decentralized Bayesian inference that characterizes scientific communication and progress. Through iterative updates and exchanges among agents, \(\boldsymbol{\theta}\) helps refine the collective understanding, driving the evolution of scientific knowledge. This dynamic interplay underscores the importance of \(\boldsymbol{\theta}\) in shaping both local and global scientific representations, ultimately contributing to the advancement of science as a collaborative endeavor.}

\begin{figure}[tb]
    \includegraphics[width=1.0\linewidth]{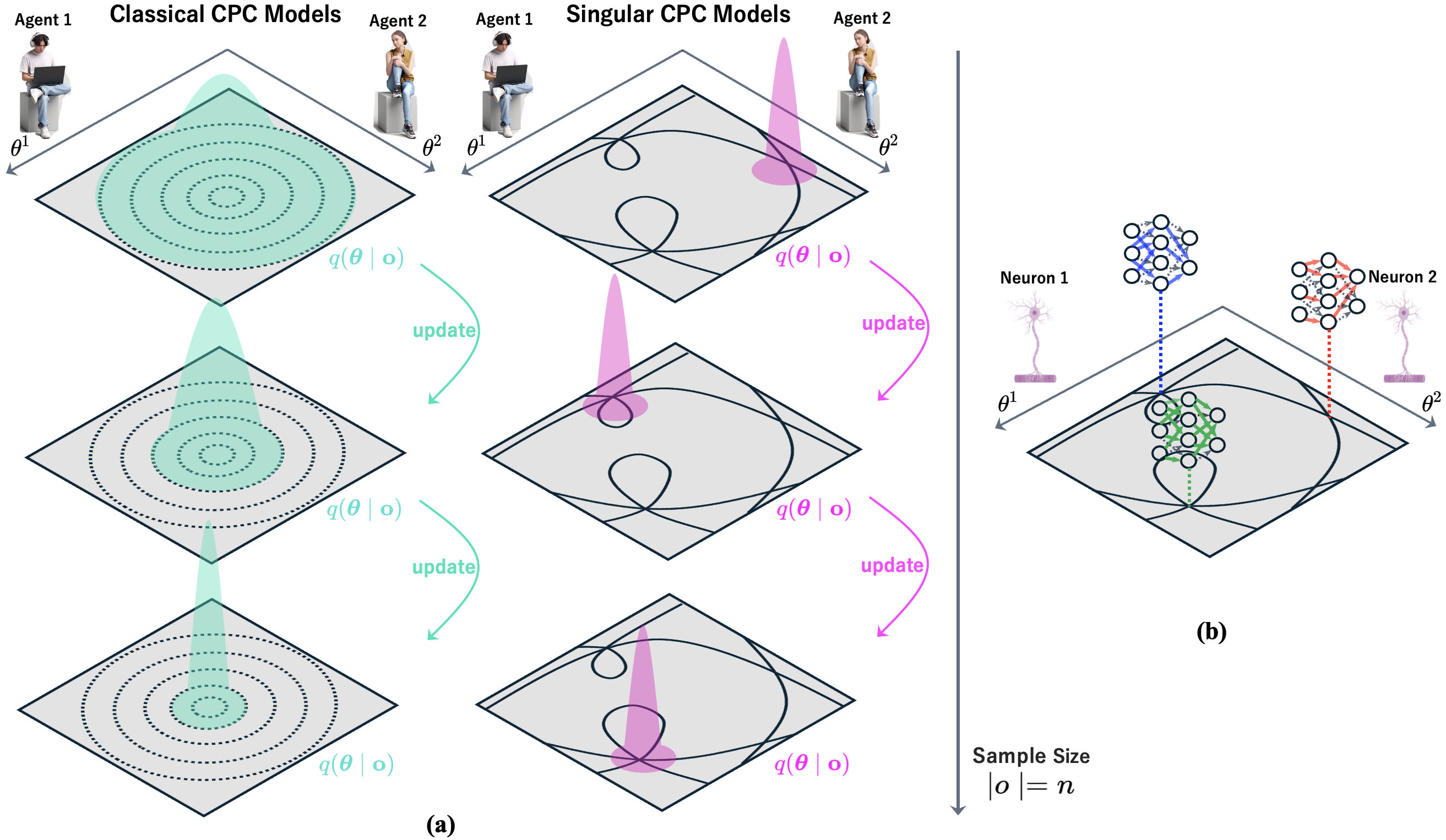}
    \caption{(a) Left: expected freecal CPC models show gradual, continuous updates of the posterior distribution $q(\boldsymbol{\theta} \mid \mathbf{o})$ as sample size increases. Right: Singular CPC models demonstrate discontinuous jumps in the posterior distribution, representing paradigm shifts in scientific understanding. (b) A network representation of the Singular CPC model, where scientists (Agent 1 and Agent 2) act as neurons. The connections between nodes illustrate how scientific concepts or theories interact and evolve, potentially leading to sudden structural changes in the collective knowledge network.}
\label{fig:singular_models}
\end{figure}

Science is expected to be progressive as well as objective, continually making solid improvements over past achievements. 
In the CPC-MS model, scientific progress is understood as the diachronic improvement of the posterior distribution through decentralized Bayesian updating. 
As scientific inquiry continues and more data $\mathbf{o}$ accumulate, the posterior distribution $q(w |\mathbf{o})$ is expected to better approximate the data-generating process. 
This can be seen as an optimization process in which probability mass is progressively concentrated around the parameters with higher likelihood (Fig.~\ref{fig:singular_models}). 
{At the same time, this is accompanied by an optimization of the model parameter \(\boldsymbol{\theta} = \left\{ \theta^k \right\}_k\), which (i) encapsulates the internal models learned by each agent, influencing how they interpret observations and form hypotheses about their environment (Fig. \ref{fig:cpc-pgm}), and (ii) acts as a bridge between individual internal representations and the shared global scientific representation \(w_d\). In this sense, \(\boldsymbol{\theta}\) can be conceived as the scientific understanding of each individual. This perspective allows us to view Bayesian optimization as a form of scientific progress---the process through which scientists' understanding of the world improves over time.}

Under this identification, the conventional Bayesian update corresponds to a gradual improvement of the posterior distribution, where the weights of neural connectivity get refined continuously and smoothly on the basis of incoming data (Fig.~\ref{fig:singular_models} (a) left). 
This mode of updating process captures the gradual scientific progress in ``normal science,'' where the scientific community as a whole concentrates on a given vicinity of the parameter space---circumscribed by one paradigm---and seeks to its refinement. 

When the parameter space of the model is singular, however, the Bayesian updating process of the posterior distribution may involve jumps from one singular point to another corresponding to a local optimal solution (Fig.~\ref{fig:singular_models} (b) right). {In singular learning theory, the concept of a singular model refers to a model whose parameter space contains singularities, as opposed to a regular model which lacks such singularities. Developed by Sumio Watanabe, this theory employs theorems from algebraic geometry to rigorously analyze the behavior of Bayesian updating in singular models\cite{thegraybook2009, thegreenbook2018}.
The key finding is that, in stark contrast to regular models, the Bayesian updating process of the posterior distribution in singular models can exhibit discontinuous jumps from one singularity to another. These jumps correspond to phase transitions between local optima and are a frequently observed phenomenon in the Bayesian updating of posterior distributions in deep learning models.
Mathematically, the singularities in the parameter space of these models determine the architecture of the model itself, dictating the presence or absence of edges connecting neurons. Consequently, learning in Bayesian deep models can be interpreted as a process whereby the model architecture is updated discontinuously via Bayesian updating of the posterior distribution.} This jump of the posterior distribution from one singularity to another is called a phase transition, a phenomenon frequently observed in the Bayesian updating process of the posterior distribution in deep learning models. In deep learning models, singularities in the parameter space correspond to the architecture of the model and determine the presence or absence of edges connecting one neuron to another. In other words, learning a Bayesian deep learning model can be interpreted as a process in which the architecture of the model is updated discontinuously through Bayesian updating of the posterior distribution. 
In the context of the CPC framework, this can be viewed as a discontinuous change in the internal representation $z$ of individual scientists, which in turn may lead to a paradigm shift in the distribution of the global latent variable $w$. 

Thomas Kuhn~\cite{Kuhn1962-wu} famously described scientific progress as consisting of two distinct periods: the period of ``normal science,'' in which scientists engage in puzzle-solving activities and piecemeal refinements of a given paradigm, and the period of ``extraordinary science,'' during which accumulated anomalies encourage the critical scrutiny of foundational assumptions and exploration of new theoretical possibilities, eventually leading to a scientific revolution. 
The CPC-MS framework offers a Bayesian model that accommodates these two periods as distinct modes of Bayesian optimization. Normal science, in this framework, is modeled by the gradual Bayesian update of the posterior distribution toward a given local optimum. During this period, scientists espouse more or less similar local representations, making proposals of external representations that deviate significantly from the shared paradigm (namely, the set of $w$'s constituting a local optimum) rare, and, if made, likely to be rejected. 
On the other hand, scientific revolutions are represented by phase transitions that leap the posterior distribution from one singularity to another in a discontinuous fashion. Such transitions induce significant reconfigurations in the collective external representations ($w$) held by the scientific community, such that a set of propositions that enjoyed high probabilities before the transition cease to do so afterward. 

Kuhn infamously claimed that the incommensurability between different paradigms makes the comparison and evaluation of competing scientific theories difficult, if not impossible, because the standards of evidence, methods, and the very concepts used in the theories can change so drastically between paradigms. 
However, within the CPC framework, there is a potential resolution to this problem that acknowledges the incommensurability while still allowing for a form of rational progress. This is because a phase transition between paradigms can still be understood as an improvement in the collective posterior distribution. 
Since a more refined posterior distribution enhances the explanatory power for observed data and sharpens the predictive accuracy for unobserved phenomena, the transition to a new paradigm can be considered a rational and objectively superior advancement in the scientific community's collective understanding, even if the paradigms themselves are internally incommensurable.
In this way, the CPC framework reconciles Kuhn's notion of incommensurability with the idea of scientific progress by proposing a unified model of Bayesian optimization that accommodates both the continuous and discontinuous phases of scientific change.

\subsection{From Confirmatory to Generative Science}
\label{section:from-confirmatory-to-generative-science}
Finally, the CPC-MS framework prompts us to reconsider the very nature and purpose of the scientific enterprise. 
The traditional realist perspective in science views its primary goal as the pursuit of true knowledge, placing strong emphasis on the confirmatory role of scientific methods to distinguish valid hypotheses from invalid ones~\cite{Carnap1950-be, Popper1959-uu}.
In the 20th century, statistical tests provided precisely such a universal tool for confirmation, establishing methodological standards across various scientific disciplines~\cite{Mayo2018-lb, Otsuka2022-eq}.
The realist picture aligns with the aforementioned image of scientists as ``lone truth seekers'' who independently collect data, test hypotheses, and publish results, the accumulation of which forms the body of scientific knowledge.

On the other hand, the CPC framework highlights the \emph{generative} aspect of science, viewing it as a coordinated system for producing new hypotheses, research plans, actions, and predictions. 
In the CPC-MS model, scientific knowledge is encoded as a posterior distribution $q(w|\mathbf{o})$ that reflects the collective opinions within of the scientific community. 
The primary function of this distributional knowledge is to induce scientists to take the next step in their inquiry. 
There are at least two such generative aspects. 
First, it serves as a key reference point for individual researchers when planning their next studies or experiments. While some researchers may strive to find exceptions to hypotheses with high posterior probability, others may design experiments that prompt a reconsideration of less attended hypotheses. In this way, scientific knowledge serves to streamline research directions that produce further evidence. This leads to a continuous cycle of proposing hypotheses, conducting experiments, and refining theories, much like the iterative updates in a Bayesian model.

The second and more obvious generative role of scientific knowledge is prediction: theories are used to predict unobserved phenomena. In the Bayesian framework, a prediction of new data $\tilde{\mathbf{o}}$ on the basis of past observation $\mathbf{o}$ is given by the posterior predictive distribution $p(\tilde{\mathbf{o}}\mid \mathbf{z}, \mathbf{a}, \mathbf{C}) q(\mathbf{z} \mid w, \mathbf{o}, \mathbf{a}) q(w \mid \mathbf{z})$
\begin{equation}
p(\tilde{\mathbf{o}}\mid \mathbf{o}, \mathbf{a}, \mathbf{C}) =  \int p(\tilde{\mathbf{o}}\mid \mathbf{z}, \mathbf{a}, \mathbf{C}) q(\mathbf{z} \mid w, \mathbf{o}, \mathbf{a}) q(w \mid \mathbf{z}) ~\mathrm{d}w \mathrm{d}\mathbf{z}.
\label{prediction}
\end{equation}
Given that an accurate posterior distribution $p(\tilde{\mathbf{o}}\mid \mathbf{o}, \mathbf{a}, \mathbf{C})$ leads to better predictions, both in terms of mean squared error and the Kullback-Leibler divergence between the predicted and true distributions, the scientific progress discussed above can also be seen as an improvement in predictive performance. The emphasis on prediction becomes even more apparent in light of the discussion on active inference (Section 2.3), where the scientific community as a whole is viewed as a collective agent aiming to minimize predictive error. 

The shift from confirmatory to generative approaches induced by the CPC framework is intrinsically linked to its social and progressive visions of science discussed above. 
From the generative perspective, scientific knowledge is seen as less an end than a means that promotes scientists to pursue new projects. Social exchanges and information sharing accelerate this research cycle, enabling the scientific community as a whole to effectively explore the vast hypothesis space in an efficient manner.
The focus on prediction also substantiates the common intuition that science is progressive, an idea that has posed a vexing issue for the confirmatory view of science---precisely because the history of science seemingly presents us with successive scrap-and-build cycles through the refutation of past theories rather than the cumulative build-up of true propositions over time~\cite{Laudan1981-ci}. The Bayesian perspective, on the other hand, offers a clear sense of scientific progress in terms of predictive performance, for the diachronic refinement of the posterior distribution, as discussed in the previous section, directly leads to the improvement of the posterior predictive distribution shown in Equation (\ref{prediction}). This suggests that the aspect of science that can be said to progress, particularly before and after a paradigm shift, is not its stock of confirmed knowledge, but rather its generative capability.

In conclusion, the CPC framework provides a comprehensive view of scientific activities that highlights their social, progressive, and generative nature. This stands in contrast to the traditional view, which portrays science as a predominantly individual, cumulative, and confirmatory procedure.
The view of science as a generative process, driven by CPC, provides a new perspective on how scientific knowledge evolves and advances. It emphasizes the importance of diversity in scientific approaches, the value of paradigm-shifting discoveries, and the collective nature of scientific progress. By understanding science through this lens, we can better appreciate the complex dynamics that drive scientific advancement and potentially develop strategies to facilitate more efficient and effective knowledge generation in the scientific community.

\section{AI and Research Automation}\label{sec:4}
As we have repeatedly emphasized throughout previous sections, CPC-MS provides a comprehensive model of scientific activities carried out by scientists. Notably, as discussed earlier, it uses probabilistic generative models to mathematically describe these processes. This unique characteristics of comprehensive modeling and probabilistic approaches enables CPC-MS to offer profound insights into: 1) the potential impact of AI on the future landscape of scientific research, and 2) the development of intuitive guidelines for implementing automated scientific activities.

\subsection{Speculating AI's Impact on Science for Shaping the Future Science}
\label{section:speculating-ais-impact-on-science-for-shaping-the-future-science}

AI has made remarkable progress, its influence spreading throughout society. Science is no exception to this trend; AI has become an innovative tool in the field, and its applications in science are rapidly expanding~\cite{xu2021artificial,wang2023scientific,zhang2023artificial}. Furthermore, AI's development isn't limited to its use as a mere tool; scientists are now exploring whether AI can do research on its own as scientist~\cite{zenil2023future,lu2024aiscientist}. They're trying to figure out if AI can come up with its own research questions, plan and run experiments, and make sense of the results. The possibility of AI acting as an autonomous scientist points towards a future novel scientific community where both human and AI scientists coexist and contribute to scientific endeavors~\cite{krenn2022scientific,messeri2024artificial}.

Given this rapid advancement of AI for science and its potential to revolutionize scientific practices, there's growing interest in how AI is affecting science. The main feature of CPC-MS is its proposal of a model of science as a CPC activity involving multiple agents. This characteristic allows us to discuss important aspects of how AI might change the nature of science. Specifically, by modeling the scientific community as a hybrid system composed of fundamentally different agents - AI and humans - CPC-MS enables us to explore how this transformation might impact science or even alter the very nature of scientific inquiry.

For instance, the entry of AI into the scientific community might reduce the overall bias of the community. Humans are constrained by biological and cognitive limitations that restrict what they can understand, and their cognition is distorted by cognitive biases and social pressures. In the CPC framework, $\theta^k$ for each agent $k$ represents these biases and hence their world models, which in turn affect the generation of $o_d^k$ and $z_d^k$. Since bias is prevalent in human agents, its effect on the global scientific representations $w_d$ can be proliferated through collective inference in human-only scientific community. Such bias would lead to skewed or biased inferences, the ``blind spot'' theory, and slower convergence, potentially degrading the social objectivity of science, as we have discussed in Section \ref{section:social-objectivity}. It could also introduce noise and bias in observations or violate the i.i.d. condition, disturbing asymptotic improvement (Section \ref{section:scientific-progress}).

On the other hand, AI would likely have a different $\theta^k$ than humans', suggesting the possibility of generating different $o_d^k$ and $z_d^k$. Thus, the entry of AI agents into the scientific community might enable the generation of $w_d$ that reflects more diverse aspects of target of study $d$. In other words, the involvement of AI in conducting science suggests that it could introduce greater diversity, which is key for social objectivity in science as discussed in Section \ref{section:social-objectivity}, into the process of scientific knowledge production. Therefore, AI potentially alleviates some of the limitations currently faced by human science, leading to potentially better scientific outcomes.

Simultaneously, the integration of AI also raises concerns about the shared common ground among social members, which is a prerequisite for objectivity as discussed in section \ref{section:social-objectivity}. CPC-MS emphasizes the role of consensus formation through communication among agents in the creation of scientific knowledge. From this perspective, science involving AI can be described as a hybrid system composed of entirely different agents - AI and humans. This bears structural similarities to the AI alignment problem~\cite{ji2023ai}. In this sense, the CPC-MS framework naturally frames the challenges of AI in science as a specific example of the broader AI alignment issue.

In current science, communication between agents works well because they are all humans who share similar biological bodies and language. However, AI lacks this commonality with us, potentially arising the issue of communication barrier. This suggests that the premise of science as distributed predictive coding, which we explained in Section \ref{section:social-objectivity}, could collapse with the participation of AI as a scientist in the scientific community. 

Moreover, a significant increase in the differences between agents could make global scientific representation production inefficient too. In a two-agent MHNG with agent $k$ and $l$, for example, agent $l$ updates $w_d$  by either accepting or rejecting proposals from agent $k$. If agent $k$ and agent $l$ are an AI and a human, respectively, their internal scientific representations $z^k_d$ and  $z^l_d$ and then the distributions they follow are expected to be significantly different. Consequently, the generated $w^k_d$ and $w^l_d$ will also be very different. Therefore, if the original $w_d$ is $w^l_d$ and the proposal is $w^k_d$ , the acceptance rate is expected to be quite low (e.g., in MHNG, $w^k_d$ is accepted with probability $\text{min}(1, \frac{P(z^k_d| \theta^k, w^l_d)}{P(z^k_d| \theta^k, w^k_d)})$, and when $w^k_d$ and $w^l_d$ differ significantly, this probability is likely to be small). This could lead to poor convergence efficiency. 

As the diversity and heterogeneity among agents increase, the common ground they share tends to decrease. This presents a fundamental trade-off: while greater diversity can enhance social objectivity in science, it also risks undermining the shared basis necessary for effective communication and collaboration. To maximize the potential of AI in expanding new scientific possibilities, it is thus crucial to proactively develop and establish this common ground among diverse constituents, in anticipation of AI scientists' future participation. This involves creating a new scientific framework that maintains the benefits of heterogeneity while ensuring effective collaboration between human and AI agents.

Like this, we can see that the integration of AI scientists into the human scientific community has the potential to significantly impact even the fundamental premises of science. Therefore the emergence of hybrid system of AI and human scientist is creating opportunities to debate and reshape the very nature of science itself, moving beyond simply considering their impact on traditional scientific practices. Humans might have only explored a limited portion of the possible space of science \cite{nielsen}, and it has been pointed out that the current scientific system faces many problems. The rise of AI offers us a chance to do more than just passively observe how science is changing. Instead, we can take an active role in reimagining and reshaping science itself.

\subsection{Guideline for Implementing Automated Total Science Activity}

CPC-MS has the potential to provide guidelines for implementing and automating the entire scientific process of a group of scientists as a machine learning model. This could bring a new perspective to the efforts of automating science, which have traditionally focused on automating specific tasks within an individual scientist's research workflow.

The pursuit of automating scientific practices by machines dates back to the early days of computing, with pioneering systems like DENDRAL~\cite{lindsay1993dendral} and BACON~\cite{langley1987scientific}. Subsequently, as exemplified by the term fourth paradigm of science, the advancement and proliferation of computational capabilities led to the automation of certain tasks within scientific activities~\cite{hey2009fourth}. The 2000s saw machine learning techniques like Bayesian optimization applied to scientific processes. However, it was the deep learning revolution of the 2010s that truly accelerated the integration of AI into science, a field now known as AI for Science~\cite{wang2023scientific,xu2021artificial,zhang2023artificial}. This integration has enabled discoveries previously unsolved by humans, with AlphaFold~\cite{jumper2021highly} serving as a prime example.

While these activities have brought tremendous progress to humanity, they have largely been limited to the automation and substitution of a specific task in a scientific process. Current research involves activities like formulating questions, conducting experiments, and writing papers.  However, attempts to fully automate an entire research cycle, encompassing all of these activities from start to finish, remain few and far between.

There are few examples of attempts to automate an entire research cycle, with systems like Adam~\cite{king2004functional} and Eve~\cite{williams2015cheaper} being notable exceptions. Adam is a closed-loop automation system that combines traditional symbolic AI with robotics to automatically generate hypotheses, verify them, and modify them based on feedback. More recently, groundbreaking research has proposed an AI system capable of conducting machine learning research in a highly automated, end-to-end manner, handling a wide range of tasks from initial idea generation to paper writing~\cite{lu2024aiscientist}.

However, even these attempts are specialized, focusing on automating specific areas of scientific research rather than achieving a general automation of science. Furthermore, the automation achieved so far typically replicates the research process of a single scientist. The AI Scientist by Lu et al. ~\cite{lu2024aiscientist} represents a significant step forward in automating the social aspects of scientific work, as it not only automates paper writing but also the peer review process. However, even this study still falls short of automating the entire scientific activities.

As this paper emphasizes, science is inherently a collaborative effort. It's common for multiple researchers to work together to understand a single phenomenon, and peer review - the process by which other researchers validate research outcomes - is a crucial aspect of scientific activity. The existing closed-loop research automation systems have not yet managed to incorporate these social aspects, leaving the automation of the entire scientific process, including its collaborative nature, still unrealized.

To achieve this goal, we need a model that captures both the social dimensions of science and the entire process from discovery to knowledge representation and communication by scientists, without being too tied to specific scientific practices. Moreover, to actually automate this model in practice, it needs to be implementable, which means it must be expressible mathematically. CPC-MS offers such a model of science. In essence, CPC-MS could serve as a foundation or starting point for developing a system capable of automatically executing the full spectrum of scientific activities performed by a scientific community.

CPC-MS takes a unique approach by modeling science as a probabilistic generative process. This perspective allows us to view the whole range of scientific activities through the lens of statistical machine learning, enabling us to potentially implement these activities as a machine learning model. Furthermore, by framing science in terms of generative models, CPC-MS can easily incorporate the latest developments in generative AI - a field that has been making waves across society in recent years. To conclude, CPC-MS offers a novel framework for understanding and advancing the future direction of research automation.

\section{Future Work}\label{sec:5}
\paragraph{Network Structure in Scientific Communities} The CPC framework can be extended by considering the network structure of multiple agents. Interactions between multiple agents with the CPC framework describe biases and effects in scientific communities. Global scientific representations are often generated by a small scientific community with a few dozen of members. Demographics of such small community may bias or influence scientific activities such as paper citation~\cite{Teich2024Citation}, peer-review\cite{Liu2023Review}, and recruiting~\cite{Lienard2018NatCommu}. For example, gendered citation patterns are known in contemporary physics~\cite{Teich2024Citation}. This pattern can be described as inferences from smaller number of agents. A small number of agents easily biases collective decision-making.

Another bias can come from whether researchers work with famous or non-famous researchers, notably the Matthew Effect ~\cite{Merton1968matthew, Bol2018matthew}. The difference between famous and less famous researchers may influence range of broadcasting global representations. Less famous researcher may not be able to broadcast global scientific representations to other members effectively while famous researcher can do. A similar pattern also appears in famous and less famous research institutes~\cite{Zhang2022advantage}. Academic success may also be influenced by structure of social network. For example, diverse intellectual synthesis between mentors and mentees influences success in academic careers~\cite{Lienard2018NatCommu}.

It is worth to mention relationship with science of science (SciSci)~\cite{Fortunato2018SciSci}. SciSci is an emergent interdisciplinary field to uncover mechanisms of doing science using computational methods. Geographical and temporal interactions between scientists are targets of SciSci in addition to conventional citation and co-authorship analyses. The CPC framework is based on interactions between multiple agents, and concurs with SciSci from the perspective of analyzing network structure.

Furthermore, SciSci also aims to influence network structure of agents and scientific activities. One solution is to create open datasets and softwares visualizing biases in scientific activities. Some studies contribute to policy making in scientific commercialization~\cite{Marx2022commercialization} and public funding~\cite{yin2022public, Ohniwa2023JapanaGrants}. Ohniwa et al., showed diverse and smaller grants are more effective than centralized and larger grants to generate new fields and promote technology transfer in Japan public funding~\cite{Ohniwa2023JapanaGrants}. Such evidence may theoretically support effectiveness of decentralized communities, so called decentralized autonomous organizations (DAOs), to support science projects \cite{Fantaccini2024DAOs}. Recently, there has been a growing momentum around the movement known as decentralized science (DeSci). DeSci are practices of science or related activities, which often are designed with decentralized mechanisms and technologies such as blockchain. In this regards, the CPC theory can be extended by theoretical understanding of social biases and computational frameworks to analyze new forms of science management models including DeSci. Further studies of social network will contribute to the extension of the CPC framework.

\paragraph{Simulation Study of CPC Science Model} CPC provides a computational model of science. Therefore, by executing this model on a computer, it is possible to simulate how the scientific community produces knowledge. The advantage of being able to simulate science is that it allows us to conduct pseudo-experiments on science, which is a social activity where actual experiments are not possible. This enables us to discuss how various factors could potentially influence the scientific community and in what ways. For instance, changing the distribution that $\theta^k$ follows for each $k$ may allow us to study how the diversity among scientists actually influences scientific activity (Sections \ref{section:social-objectivity} and \ref{section:speculating-ais-impact-on-science-for-shaping-the-future-science}). Moreover, if we limit the communication to agents within the same cluster $c$ and restrict inter-cluster communication, we may be able to reveal the impact of clustering on the production of global scientific representations in scientific communities, as discussed above. Additionally, introducing malicious players who make false reports. For example, we may introduce players with an irrational acceptance ratio of $w_d$ or players that randomly suggest $w^k_d$ independent of $z^k_d$. Alternatively we may also introduce agents that excessively prioritize individual epistemic value or pragmatic values (Section~\ref{sec:2.3}) and how these agents affect scientific integrity.

\paragraph{Non-Statioanrity} In the current CPC-MS, for simplicity, it is implicitly assumed that the distribution that $w_d$ follows is stationary. However, in reality, the distribution that $w_d$ follows may change over time. For example, in fields where claims are difficult to verify and established theories can change significantly when new evidence emerges, the distribution of $w_d$ is expected to be non-stationary. Extending CPC-MS to naturally model such non-stationary distributions of $w_d$ is one potential direction for expanding the framework.

\paragraph{Incorporating External Players}
In the current framework, only scientists are considered agents, but in actual science, multiple non-scientist players are essential. These include funders, influencers of scientific direction, and educators. The current CPC-MS framework does not explicitly model these players, limiting discussions of their influence. Incorporating non-scientist players could extend the model. For example, funders influence research themes, indirectly corresponding to selecting certain object $d$ to be studied from this world.  This decision-making is closely related to how $w_d$ is used and what is required as $w_d$ for a society~\cite{yin2022public}. The current framework models the inference of $w_d$ of target of study but does not address how the target is selected or how $w_d$ is used. Modeling such target selection process, as mentioned above, could provide deeper insights into the effect of the society to science.

\paragraph{Utilities and Values}
When selecting research topics, the value to society and individuals becomes a crucial factor driving research activities. At first glance, the generative model of CPC (Figure~\ref{fig:cpc-pgm}) may appear to focus solely on the encoding of observational information, without incorporating elements such as individual values or rewards. However, the PGM of CPC can be naturally extended to include decision-making based on rewards and values. In many cases, decision-making based on rewards and values can be reformulated as Bayesian inference of action sequences to reach a desired state~\cite{levine2018reinforcement,kappen2012optimal}. Building on this theory, Ebara et al. have proposed a multi-agent reinforcement learning algorithm that extends the MHNG based on CPC to include symbol emergence~\cite{ebara2023multi}.

Following the formulation of active inference presented in Section~\ref{sec:2.3}, pragmatic value can be used to model both utilities and values.

 Based on the CPC-MS framework, future challenges include discussing the selection of research topics based on values and the impact of individual agents' (researchers') incentives on scientific exploration as CPC.




\section{Conclusion}\label{sec:6}
The CPC-MS framework presented in this paper offers a novel perspective on scientific activities, viewing them through the lens of collective predictive coding and decentralized Bayesian inference. By modeling science as a generative process carried out by a community of agents—namely, generative science—CPC-MS provides several key insights.

It formalizes the social nature of scientific knowledge production, demonstrating how individual observations and hypotheses are integrated into explicit scientific knowledge, i.e., global scientific representations, through communication and peer review. The CPC-MS framework offers a mathematical foundation for understanding scientific progress, paradigm shifts, and the role of diversity in scientific communities. It bridges the gap between individual cognitive processes and collective knowledge creation in science, offering a unified view of scientific activities, from experimentation to theory development.

The CPC-MS framework aligns with and extends existing ideas in the philosophy of science, such as social objectivity and the generative nature of scientific theories. It encourages us to view science not just as a collection of facts or theories, but as a dynamic, collective cognitive process that continually refines our understanding of the world.

As we move towards an era where AI plays an increasingly significant role in scientific research, the CPC-MS framework provides a valuable tool for understanding and shaping the future of science. Additionally, the mathematical nature of CPC-MS allows us to integrate generative AI-based automated scientific research as a part of the CPC-MS framework.

Future work could focus on refining the mathematical models underlying CPC-MS, conducting empirical studies to test its validity in modeling actual scientific studies, and exploring its implications for science policy, research management, and research ethics. Ultimately, this framework aims to contribute to a more comprehensive understanding of how scientific knowledge is created, validated, and advanced through collective effort.

\section*{Acknowledgment}
This study is partially supported by JSPS KAKENHI Grant Number JP21H04904, and JST Moonshot R\&D Program, Grant Number JPMJMS2033.
We also thank Ryuichi Maruyama and the AI Alignment Network (ALIGN), who facilitated our collaboration and provided continuous support.

\bibliographystyle{alpha}
\bibliography{main}

\newcommand{\etalchar}[1]{$^{#1}$}
\begin{thebibliography}{YDW{\etalchar{+}}22}

\bibitem[Bak16]{baker20161}
Monya Baker.
\newblock 1,500 scientists lift the lid on reproducibility.
\newblock {\em Nature}, 533(7604), 2016.

\bibitem[Ber50]{Berlyne1950curiosity}
Daniel Berlyne.
\newblock Novelty and curiosity as determinants of exploratory behavior.
\newblock {\em British Journal of Psychology}, 41:68--80, 1950.

\bibitem[BfCF{\etalchar{+}}23]{balzan2023distributed}
Francesco Balzan-francesco, John Campbell, Karl Friston, Maxwell~James Ramstead, Daniel Friedman, and Axel Constant.
\newblock Distributed science-the scientific process as multi-scale active inference.
\newblock {\em OSF Preprints}, 2023.

\bibitem[Bis06]{bishop2006pattern}
Christopher~M Bishop.
\newblock {\em Pattern recognition and machine learning}.
\newblock springer, 2006.

\bibitem[Car50]{Carnap1950-be}
Rudolf Carnap.
\newblock {\em Logical foundations of probability}.
\newblock University of Chicago Press, 1950.

\bibitem[Cha02]{Chandler2002}
Daniel Chandler.
\newblock {\em {Semiotics the Basics}}.
\newblock Routledge, 2002.

\bibitem[Cla13]{clark2013whatever}
Andy Clark.
\newblock Whatever next? predictive brains, situated agents, and the future of cognitive science.
\newblock {\em Behavioral and brain sciences}, 36(3):181--204, 2013.

\bibitem[ENTT23]{ebara2023multi}
Hiroto Ebara, Tomoaki Nakamura, Akira Taniguchi, and Tadahiro Taniguchi.
\newblock Multi-agent reinforcement learning with emergent communication using discrete and indifferentiable message.
\newblock In {\em 2023 15th International Congress on Advanced Applied Informatics Winter (IIAI-AAI-Winter)}, pages 366--371, 2023.

\bibitem[FBB{\etalchar{+}}18]{Fortunato2018SciSci}
Santo Fortunato, Carl~T. Bergstrom, Katy Börner, James~A. Evans, Dirk Helbing, Staša Milojević, Alexander~M. Petersen, Filippo Radicchi, Roberta Sinatra, Brian Uzzi, Alessandro Vespignani, Ludo Waltman, Dashun Wang, and Albert-László Barabási.
\newblock Science of science.
\newblock {\em Science}, 359(6379):eaao0185, 2018.

\bibitem[FMN{\etalchar{+}}21]{friston2021world}
Karl Friston, Rosalyn~J Moran, Yukie Nagai, Tadahiro Taniguchi, Hiroaki Gomi, and Josh Tenenbaum.
\newblock World model learning and inference.
\newblock {\em Neural Networks}, 144:573--590, 2021.

\bibitem[Fri10]{friston2010free}
Karl Friston.
\newblock The free-energy principle: a unified brain theory?
\newblock {\em Nature reviews neuroscience}, 11(2):127--138, 2010.

\bibitem[Fri19]{friston2019free}
Karl Friston.
\newblock A free energy principle for a particular physics.
\newblock {\em arXiv preprint arXiv:1906.10184}, 2019.

\bibitem[HDW{\etalchar{+}}23]{hope2023computational}
Tom Hope, Doug Downey, Daniel~S Weld, Oren Etzioni, and Eric Horvitz.
\newblock A computational inflection for scientific discovery.
\newblock {\em Communications of the ACM}, 66(8):62--73, 2023.

\bibitem[Hey09]{hey2009fourth}
Tony Hey.
\newblock {\em The fourth paradigm}.
\newblock United States of America., 2009.

\bibitem[Hoh13]{hohwy2013predictive}
Jakob Hohwy.
\newblock {\em The predictive mind}.
\newblock Oxford University Press, 2013.

\bibitem[HS18]{ha2018world}
David Ha and J{\"u}rgen Schmidhuber.
\newblock World models.
\newblock {\em arXiv preprint arXiv:1803.10122}, 2018.

\bibitem[JEP{\etalchar{+}}21]{jumper2021highly}
John Jumper, Richard Evans, Alexander Pritzel, Tim Green, Michael Figurnov, Olaf Ronneberger, Kathryn Tunyasuvunakool, Russ Bates, Augustin {\v{Z}}{\'\i}dek, Anna Potapenko, et~al.
\newblock Highly accurate protein structure prediction with alphafold.
\newblock {\em nature}, 596(7873):583--589, 2021.

\bibitem[JQC{\etalchar{+}}23]{ji2023ai}
Jiaming Ji, Tianyi Qiu, Boyuan Chen, Borong Zhang, Hantao Lou, Kaile Wang, Yawen Duan, Zhonghao He, Jiayi Zhou, Zhaowei Zhang, et~al.
\newblock Ai alignment: A comprehensive survey.
\newblock {\em arXiv preprint arXiv:2310.19852}, 2023.

\bibitem[KGO12]{kappen2012optimal}
Hilbert~J Kappen, Vicen{\c{c}} G{\'o}mez, and Manfred Opper.
\newblock Optimal control as a graphical model inference problem.
\newblock {\em Machine learning}, 87(2):159--182, 2012.

\bibitem[KHK{\etalchar{+}}09]{Kang2009IG}
Min~Jeong Kang, Ming Hsu, Ian~M Krajbich, George Loewenstein, Samuel~M McClure, Joseph Tao-yi Wang, and Colin~F Camerer.
\newblock The wick in the candle of learning: epistemic curiosity activates reward circuitry and enhances memory.
\newblock {\em Psychological Research}, 20(9):963--73, 2009.

\bibitem[Kit93]{Kitcher1993-ly}
Philip Kitcher.
\newblock {\em The Advancement of Science: Science Without Legend, Objectivity Without Illusions}.
\newblock Oxford University Press, 1993.

\bibitem[KPG{\etalchar{+}}22]{krenn2022scientific}
Mario Krenn, Robert Pollice, Si~Yue Guo, Matteo Aldeghi, Alba Cervera-Lierta, Pascal Friederich, Gabriel dos Passos~Gomes, Florian H{\"a}se, Adrian Jinich, AkshatKumar Nigam, et~al.
\newblock On scientific understanding with artificial intelligence.
\newblock {\em Nature Reviews Physics}, 4(12):761--769, 2022.

\bibitem[KRO{\etalchar{+}}09]{king2009automation}
Ross~D King, Jem Rowland, Stephen~G Oliver, Michael Young, Wayne Aubrey, Emma Byrne, Maria Liakata, Magdalena Markham, Pinar Pir, Larisa~N Soldatova, et~al.
\newblock The automation of science.
\newblock {\em Science}, 324(5923):85--89, 2009.

\bibitem[Kuh62]{Kuhn1962-wu}
Thomas~S Kuhn.
\newblock {The Structure of Scientific Revolutions}.
\newblock {\em University of Chicago Press}, page 226, 1962.

\bibitem[KWJ{\etalchar{+}}04]{king2004functional}
Ross~D King, Kenneth~E Whelan, Ffion~M Jones, Philip~GK Reiser, Christopher~H Bryant, Stephen~H Muggleton, Douglas~B Kell, and Stephen~G Oliver.
\newblock Functional genomic hypothesis generation and experimentation by a robot scientist.
\newblock {\em Nature}, 427(6971):247--252, 2004.

\bibitem[LAAD18]{Lienard2018NatCommu}
Jean~F. Liénard, Titipat Achakulvisut, Daniel~E. Acuna, and Stephen~V. David.
\newblock Intellectual synthesis in mentorship determines success in academic careers.
\newblock {\em Nature Communications}, 9:4840, 2018.

\bibitem[Lak78]{Lakatos1978-fw}
Imre Lakatos.
\newblock {\em {The methodology of scientific research programmes}}.
\newblock Cambridge University Press, 1978.

\bibitem[Lan87]{langley1987scientific}
Pat Langley.
\newblock {\em Scientific discovery: Computational explorations of the creative processes}.
\newblock MIT press, 1987.

\bibitem[Lau81]{Laudan1981-ci}
Larry Laudan.
\newblock A confutation of convergent realism.
\newblock {\em Philos. Sci.}, 48(1):19--49, March 1981.

\bibitem[LBFL93]{lindsay1993dendral}
Robert~K Lindsay, Bruce~G Buchanan, Edward~A Feigenbaum, and Joshua Lederberg.
\newblock Dendral: a case study of the first expert system for scientific hypothesis formation.
\newblock {\em Artificial intelligence}, 61(2):209--261, 1993.

\bibitem[Lev18]{levine2018reinforcement}
Sergey Levine.
\newblock Reinforcement learning and control as probabilistic inference: Tutorial and review.
\newblock {\em arXiv preprint arXiv:1805.00909}, 2018.

\bibitem[LL82]{Landau1982}
L.D. Landau and E.M. Lifshitz.
\newblock {\em Mechanics: Volume 1}.
\newblock Number v. 1. Elsevier Science, 1982.

\bibitem[LLL{\etalchar{+}}24]{lu2024aiscientist}
Chris Lu, Cong Lu, Robert Lange, Jakob~N Foerster, Jeff Clune, and David Ha.
\newblock The ai scientist: Towards fully automated open-ended scientific discovery.
\newblock {\em arXiv preprint arXiv:2408.06292}, 2024.

\bibitem[Loe94]{Lowewenstein1994IG}
George Loewenstein.
\newblock The psychology of curiosity: A review and reinterpretation.
\newblock {\em Psychological Bulletin}, 116(1):75--98, 1994.

\bibitem[Lon90]{Longino1990-ek}
Helen~E Longino.
\newblock {\em Science as social knowledge: Values and objectivity in scientific inquiry}.
\newblock Princeton University Press, Princeton, NJ, 1990.

\bibitem[LRA18]{Liu2023Review}
Fengyuan Liu, Talal Rahwan, and Bedoor AlShebli.
\newblock Non-white scientists appear on fewer editorial boards, spend more time under review, and receive fewer citations.
\newblock {\em Proceedings of the National Academy of Sciences}, 9:4840, 2018.

\bibitem[LW79]{latour1979laboratory}
Bruno Latour and Steve Woolgar.
\newblock {\em Laboratory life: The construction of scientific facts}.
\newblock Sage Publications, 1979.

\bibitem[LZC{\etalchar{+}}24]{liang2024can}
Weixin Liang, Yuhui Zhang, Hancheng Cao, Binglu Wang, Daisy~Yi Ding, Xinyu Yang, Kailas Vodrahalli, Siyu He, Daniel~Scott Smith, Yian Yin, et~al.
\newblock Can large language models provide useful feedback on research papers? a large-scale empirical analysis.
\newblock {\em NEJM AI}, page AIoa2400196, 2024.

\bibitem[LZW{\etalchar{+}}24]{liang2024mapping}
Weixin Liang, Yaohui Zhang, Zhengxuan Wu, Haley Lepp, Wenlong Ji, Xuandong Zhao, Hancheng Cao, Sheng Liu, Siyu He, Zhi Huang, et~al.
\newblock Mapping the increasing use of llms in scientific papers.
\newblock {\em arXiv preprint arXiv:2404.01268}, 2024.

\bibitem[May18]{Mayo2018-lb}
Deborah~G Mayo.
\newblock {\em {Statistical Inference as Severe Testing: How to Get Beyond the Statistics Wars}}.
\newblock Cambridge University Press, Cambridge, UK, 2018.

\bibitem[MC24]{messeri2024artificial}
Lisa Messeri and MJ~Crockett.
\newblock Artificial intelligence and illusions of understanding in scientific research.
\newblock {\em Nature}, 627(8002):49--58, 2024.

\bibitem[Mer68]{Merton1968matthew}
Robert~K. Merton.
\newblock The matthew effect in science: The reward and communication systems of science are considered.
\newblock {\em Science}, 159(3810):56--63, 1968.

\bibitem[MH18]{Bol2018matthew}
Matt Marx and David~H Hsu.
\newblock The matthew effect in science funding.
\newblock {\em Proceedings of the National Academy of Sciences}, 115(19):4887--4890, 2018.

\bibitem[MH21]{Marx2022commercialization}
Matt Marx and David~H Hsu.
\newblock Revisiting the entrepreneurial commercialization of academic science: Evidence from “twin” discoveries.
\newblock {\em Management Science}, 68(2):1330--1352, 2021.

\bibitem[NQ22]{nielsen}
Michael Nielsen and Kanjun Qiu.
\newblock {A Vision of Metascience An Engine of Improvement for the Social Processes of Science}.
\newblock \url{https://scienceplusplus.org/metascience/}, 2022.
\newblock Accessed on 2023-05-17.

\bibitem[OGL16]{Oudeyer2016LP}
Pierre Oudeyer, Jacqueline Gottlieb, and Manuel Lopes.
\newblock Intrinsic motivation, curiosity, and learning: Theory and applications in educational technologies.
\newblock {\em Progress in Brain Research}, 229:257--284, 2016.

\bibitem[Ots22]{Otsuka2022-eq}
Jun Otsuka.
\newblock {\em Thinking About Statistics: The Philosophical Foundations}.
\newblock Routledge, December 2022.

\bibitem[Pop59]{Popper1959-uu}
Karl~Raimund Popper.
\newblock {\em {The Logic of Scientific Discovery}}.
\newblock Hutchinson \& Co, 1959.

\bibitem[RTH23]{Ohniwa2023JapanaGrants}
Ohniwa~L Ryosuke, Kunio Takeyasu, and Aiko Hibino.
\newblock The effectiveness of japanese public funding to generate emerging topics in life science and medicine.
\newblock {\em PLOS ONE}, 18(8):e0290077, 2023.

\bibitem[SJNK24]{Spitzer2024IG}
W.~H.~Markus Spitzer, Jania Janz, Maohua Nie, and Andrea Kiesel.
\newblock On the interplay of curiosity, confdence, and importance in knowing information.
\newblock {\em Psychological Research}, 88:101--115, 2024.

\bibitem[SLA24]{Fantaccini2024DAOs}
Fantaccini Simone, Grassi Laura, and Rampoldi Andrea.
\newblock The potential of daos for funding and collaborative development in the life sciences.
\newblock {\em Nat Biotechnol}, 42:555--562, 2024.

\bibitem[SPH{\etalchar{+}}19]{Schwartenbeck2019exploration}
Philipp Schwartenbeck, Johannes Passecker, Tobias~U Hauser, Thomas~HB FitzGerald, Martin Kronbichler, and Karl~J Friston.
\newblock Computational mechanisms of curiosity and goal-directed exploration.
\newblock {\em eLife}, 8:e41703, 2019.

\bibitem[Tan24]{taniguchi2024collective}
Tadahiro Taniguchi.
\newblock Collective predictive coding hypothesis: Symbol emergence as decentralized {Bayesian} inference.
\newblock {\em Frontiers in Robotics and AI}, 11:1353870, 2024.

\bibitem[TKL{\etalchar{+}}22]{Teich2024Citation}
Erin~G. Teich, Jason~Z. Kim, Christopher~W. Lynn, Samantha~C. Simon, Andrei~A. Klishin, Karol~P. Szymula, Pragya Srivastava, Lee~C. Bassett, Perry Zurn, Jordan~D. Dworkin, and Dani~S. Bassett.
\newblock Citation inequity and gendered citation practices in contemporary physics.
\newblock {\em Nature Physics}, 6(10):1161--1170, 2022.

\bibitem[TMS{\etalchar{+}}23]{taniguchi2023world}
Tadahiro Taniguchi, Shingo Murata, Masahiro Suzuki, Dimitri Ognibene, Pablo Lanillos, Emre Ugur, Lorenzo Jamone, Tomoaki Nakamura, Alejandra Ciria, Bruno Lara, and Giovanni Pezzulo.
\newblock World models and predictive coding for cognitive and developmental robotics: frontiers and challenges.
\newblock {\em Advanced Robotics}, 37(13):780--806, 2023.

\bibitem[TNN{\etalchar{+}}16]{taniguchi2016symbol}
Tadahiro Taniguchi, Takayuki Nagai, Tomoaki Nakamura, Naoto Iwahashi, Tetsuya Ogata, and Hideki Asoh.
\newblock Symbol emergence in robotics: a survey.
\newblock {\em Advanced Robotics}, 30(11-12):706--728, 2016.

\bibitem[TP22]{parr2022activeinference}
Karl J.~Friston Thomas~Parr, Giovanni~Pezzulo.
\newblock {\em Active Inference: The Free Energy Principle in Mind, Brain, and Behavior}.
\newblock The MIT Press, 2022.

\bibitem[TUH{\etalchar{+}}18]{taniguchi2018symbol}
Tadahiro Taniguchi, Emre Ugur, Matej Hoffmann, Lorenzo Jamone, Takayuki Nagai, Benjamin Rosman, Toshihiko Matsuka, Naoto Iwahashi, Erhan Oztop, Justus Piater, et~al.
\newblock Symbol emergence in cognitive developmental systems: a survey.
\newblock {\em IEEE Transactions on Cognitive and Developmental Systems}, 11(4):494--516, 2018.

\bibitem[TYM{\etalchar{+}}23]{taniguchi2022emergent}
Tadahiro Taniguchi, Yuto Yoshida, Yuta Matsui, Nguyen~Le Hoang, Akira Taniguchi, and Yoshinobu Hagiwara.
\newblock Emergent communication through metropolis-hastings naming game with deep generative models.
\newblock {\em Advanced Robotics}, 37(19):1266--1282, 2023.

\bibitem[Wat09]{thegraybook2009}
Sumio Watanabe.
\newblock {\em Algebraic Geometry and Statistical Learning Theory}.
\newblock Cambridge University Press, 2009.

\bibitem[Wat18]{thegreenbook2018}
Sumio Watanabe.
\newblock {\em Mathematical Theory of Bayesian Statistics}.
\newblock Routledge, 2018.

\bibitem[WBS{\etalchar{+}}15]{williams2015cheaper}
Kevin Williams, Elizabeth Bilsland, Andrew Sparkes, Wayne Aubrey, Michael Young, Larisa~N Soldatova, Kurt De~Grave, Jan Ramon, Michaela De~Clare, Worachart Sirawaraporn, et~al.
\newblock Cheaper faster drug development validated by the repositioning of drugs against neglected tropical diseases.
\newblock {\em Journal of the Royal society Interface}, 12(104):20141289, 2015.

\bibitem[WFD{\etalchar{+}}23]{wang2023scientific}
Hanchen Wang, Tianfan Fu, Yuanqi Du, Wenhao Gao, Kexin Huang, Ziming Liu, Payal Chandak, Shengchao Liu, Peter Van~Katwyk, Andreea Deac, et~al.
\newblock Scientific discovery in the age of artificial intelligence.
\newblock {\em Nature}, 620(7972):47--60, 2023.

\bibitem[XLC{\etalchar{+}}21]{xu2021artificial}
Yongjun Xu, Xin Liu, Xin Cao, Changping Huang, Enke Liu, Sen Qian, Xingchen Liu, Yanjun Wu, Fengliang Dong, Cheng-Wei Qiu, et~al.
\newblock Artificial intelligence: A powerful paradigm for scientific research.
\newblock {\em The Innovation}, 2(4), 2021.

\bibitem[YDW{\etalchar{+}}22]{yin2022public}
Yian Yin, Yuxiao Dong, Kuansan Wang, Dashun Wang, and Benjamin~F Jones.
\newblock Public use and public funding of science.
\newblock {\em Nature human behaviour}, 6(10):1344--1350, 2022.

\bibitem[ZTA{\etalchar{+}}23]{zenil2023future}
Hector Zenil, Jesper Tegn{\'e}r, Felipe~S Abrah{\~a}o, Alexander Lavin, Vipin Kumar, Jeremy~G Frey, Adrian Weller, Larisa Soldatova, Alan~R Bundy, Nicholas~R Jennings, et~al.
\newblock The future of fundamental science led by generative closed-loop artificial intelligence.
\newblock {\em arXiv preprint arXiv:2307.07522}, 2023.

\bibitem[ZWH{\etalchar{+}}23]{zhang2023artificial}
Xuan Zhang, Limei Wang, Jacob Helwig, Youzhi Luo, Cong Fu, Yaochen Xie, Meng Liu, Yuchao Lin, Zhao Xu, Keqiang Yan, et~al.
\newblock Artificial intelligence for science in quantum, atomistic, and continuum systems.
\newblock {\em arXiv preprint arXiv:2307.08423}, 2023.

\bibitem[ZWLC22]{Zhang2022advantage}
Sam Zhang, K.~Hunter Wapman, Daniel~B. Larremore, and Aaron Clauset.
\newblock Labor advantages drive the greater productivity of faculty at elite universities.
\newblock {\em Science Advances}, 8(46):eabq7056, 2022.

\end{thebibliography}

\appendix

\section{Nomenclature}\label{sec:nomenclature}

\begin{table}[ht]
\centering
\caption{Nomenclature of CPC-MS}
\begin{tabular}{|c|l|}
\hline
\textbf{Symbol} & \textbf{Description} \\
\hline
$w_d$ & External representation for the $d$-th target \\
$z^k_d$ & Internal representation of the $k$-th agent for the $d$-th target \\
$o^k_d$ & Observation of the $k$-th agent for the $d$-th target \\
$d$ & Index for taret objects or phenomena being studied \\
$k$ & Index for individual agents (e.g., scientists) \\
$\theta^k$ & Local parameters for the $k$-th agent (e.g., learned internal models) \\
$K$ & Total number of agents \\
$D$ & Total number of objects \\
$p(w_d)$ & Prior distribution of external representations for the $d$-th object \\
$p(o^k_d|z^k_d)$ & Likelihood of observations given internal representations for the $k$-th agent and $d$-th object \\
$p(z^k_d|w_d)$ & Prior distribution of internal representations given external representations \\
$q(w_d|\{z^k_d\}_k)$ & Approximate posterior distribution of external given internal representations \\
$q(z^k_d|o^k_d)$ & Approximate posterior distribution of internal representations given observations \\
$p(w_d|\{o^k_d\}_k)$ & True posterior distribution of external representations given observations \\
$q(w_d|\{o^k_d\}_k)$ & Approximated posterior distribution of external representations given observations \\
$a^k$ & Actions taken by the $k$-th agent (e.g., conducting experiments) \\
$C^k$ & Rewards taken by the $k$-th agent (e.g., budget for the experiments) \\
$F$ & Variational free energy \\
$G(\mathbf{a})$ & Expected free energy for future actions (bold $\mathbf{a}$ represents actions of all agents) \\
\hline
\end{tabular}
\end{table}

\end{document}